\documentclass[fleqn,usenatbib,useAMS]{mnras}
\usepackage{graphicx}	
\usepackage{amsmath}	
\usepackage{amssymb}	
\usepackage{multicol}       
\usepackage{bm}		
\usepackage{pdflscape}
\usepackage[version=4]{mhchem}
\usepackage{adjustbox}
\usepackage[dvipsnames]{xcolor}

\usepackage[T1]{fontenc}
\usepackage{ae,aecompl}
\usepackage{newtxtext,newtxmath}

\title[Ethylene glycol towards G358.93--0.03 MM1]{Detection of antifreeze molecule ethylene glycol in the hot molecular core G358.93--0.03 MM1}

\author[Manna, Pal \& Viti]{Arijit Manna$^{1}$\thanks{E-mail:amanna.astro@gmail.com}, Sabyasachi Pal$^{1}$\thanks{E-mail:sabya.pal@gmail.com}, Serena Viti$^{2,3}$\\
	$^{1}$Department of Physics and Astronomy, Midnapore City College, Paschim Medinipur, West Bengal, India 721129\\
	$^{2}$Leiden Observatory, Leiden University, PO Box 9513, 2300 RA Leiden, The Netherlands\\
	$^{3}$Department of Physics and Astronomy, UCL, Gower Street, London WC1E 6BT, UK
}

\date{}

\pubyear{2024}
\begin{document}
\label{firstpage}
\pagerange{\pageref{firstpage}--\pageref{lastpage}}
\maketitle

\begin{abstract}
The identification of complex prebiotic molecules using millimeter and submillimeter telescopes allows us to understand how the basic building blocks of life are formed in the universe. In the interstellar medium (ISM), ethylene glycol ((\ce{CH2OH})$_{2}$) is the simplest sugar alcohol molecule, and it is the reduced alcohol of the simplest sugar-like molecule, glycolaldehyde (\ce{CH2OHCHO}). We present the first detection of the rotational emission lines of $aGg^{\prime}$ conformer of ethylene glycol ((\ce{CH2OH})$_{2}$) towards the hot molecular core G358.93--0.03 MM1 using the Atacama Large Millimeter/Submillimeter Array (ALMA). The estimated column density of $aGg^{\prime}$-(\ce{CH2OH})$_{2}$ towards the G358.93--0.03 MM1 is (4.5$\pm$0.1)$\times$10$^{16}$ cm$^{-2}$ with an excitation temperature of 155$\pm$35 K. The abundance of $aGg^{\prime}$-(\ce{CH2OH})$_{2}$ with respect to \ce{H2} is (1.4$\pm$0.5)$\times$10$^{-8}$. Similarly, the abundances of $aGg^{\prime}$-(\ce{CH2OH})$_{2}$ with respect to \ce{CH2OHCHO} and \ce{CH3OH} are 3.1$\pm$0.5 and (6.1$\pm$0.3)$\times$10$^{-3}$. We compare the estimated abundance of $aGg^{\prime}$-(\ce{CH2OH})$_{2}$ with the existing three-phase warm-up chemical model abundance of (\ce{CH2OH})$_{2}$, and we notice the observed abundance and modelled abundance are nearly similar. We discuss the possible formation pathways of $aGg^{\prime}$-(\ce{CH2OH})$_{2}$ towards the hot molecular cores, and we find that $aGg^{\prime}$-(\ce{CH2OH})$_{2}$ is probably created via the recombination of two \ce{CH2OH} radicals on the grain surface of G358.93--0.03 MM1.
\end{abstract}

\begin{keywords}
ISM: individual objects (G358.93--0.03 MM1) -- ISM: abundances -- ISM: kinematics and dynamics -- stars:
formation -- astrochemistry
\end{keywords}

\begingroup
\let\clearpage\relax
%\tableofcontents
\endgroup
\newpage

\section{Introduction}
\label{sec:intro}
In the past few years, the higher sensitivity of radio and (sub)millimeter telescopes have made it possible to identify simple and complex molecules with an increasing number of atoms in the interstellar medium (ISM). Interstellar molecules, which contain more than six atoms, including carbon, are known as complex organic molecules (hereafter COMs) \citep{van98, her09}. COMs are mainly found in dense collapsing cloud cores \citep{bac12, taq17}, protoplanetary disks \citep{wal16, fav18}, hot molecular cores (hereafter HMCs) \citep{riv17, jo20, mondal21}, high-mass protostars \citep{is13, man24a}, and hot corinos \citep{bl87, fa15, ber17, ce17, os18}. The presence of complex prebiotic molecules in the ISM suggests that those molecules are created during the initial stage of star formation and that they are preserved until the development of small bodies. The formation mechanisms of COMs in the star-forming regions are being intensively debated in astrochemistry. Two possible routes have been proposed to create COMs: gas phase and grain surface chemical reactions \citep{vi04, gar06, gar08, gar13, ba15, sk18, gar22, en22, puz22, ce23}. Only the identification of COMs and their relative abundances and two- or three-phase warm-up chemical models in a large number of star-forming regions and HMCs will help us understand the chemical formation pathways of the COMs \citep{vi04, gar13, cou18, gar22}.

\begin{figure}
\centering
\includegraphics[width=0.47\textwidth]{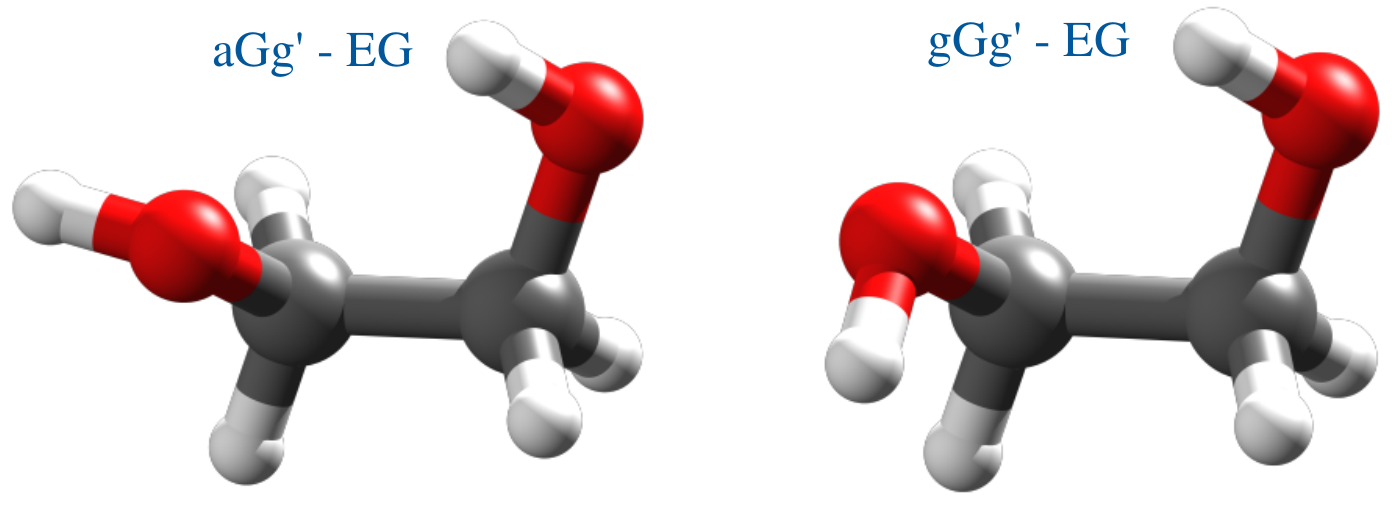}
\caption{Three-dimensional molecular structure of $aGg^{\prime}$-EG and $gGg^{\prime}$-EG. The grey atoms are carbon (C), the red atoms are oxygen (O), and the white atoms are hydrogen (H).}
\label{fig:molecule}
\end{figure}

Ethylene glycol (hereafter EG) is known as a di-alcohol molecule that is commonly used in prebiotic sugar synthesis. The EG molecule is the sugar alcohol of the aldehyde sugar molecule glycolaldehyde (hereafter GA). EG is an asymmetric top molecule with coupled rotation around its two C--O bonds and one C--C bond, which generates different conformers \citep{ch95, ch01}. There are six conformers with respect to the C--C bond with gauche (G) arrangement of the two \ce{H2C} = \ce{CH2} groups and four low-stable conformers with anti-arrangement of the \ce{CH2} groups. In the G conformer, the two OH groups pick the gauche orientation with respect to each other. The `a' and `g' notations are used for the anti (a) and gauche (g) conformers of the OH groups with respect to the two C--O bonds. When the OH group provides the hydrogen for the intramolecular bonding, the gauche orientation is noted as $aGg^{\prime}$. The EG molecule has only two conformers, $aGg^{\prime}$ and $gGg^{\prime}$. The 3D molecular structure of $aGg^{\prime}$-EG and $gGg^{\prime}$-EG is shown in Fig.~\ref{fig:molecule}. The $gGg^{\prime}$ conformer of EG is 200 cm$^{-1}$ or 290 K higher in energy than $aGg^{\prime}$ \citep{mu04}. For $aGg^{\prime}$ and $gGg^{\prime}$ conformers of EG, the tunneling is observed between two equivalent equilibrium configurations, and it splits each rotational level into two different states, $v$ = 0 and $v$ = 1 \citep{ch01}. For the $aGg^{\prime}$ and $gGg^{\prime}$ conformers, the $v$ = 1 state is around 7 and 1.4 GHz higher than the $v$ = 0 \citep{mu04}. The emission lines of EG were first detected towards the galactic center along the line of sight to Sgr B2, but the several detected lines of EG were blended \citep{hol02, req08, bel13}. Evidence of EG was also found towards the HMCs Orion KL, W51 e2, G34.3+0.2, G31.41+0.31, G10.47+0.03 \citep{fav11, fu14, bro15, ly15, riv17, mondal21}, and hot corinos IRAS 16293-2422 B, NGC 1333 IRAS 2A \citep{jo16, cou15}. Additionally, the emission lines of EG were also found towards the comets Hale-Bopp, C/2012 F6 (Lemmon), C/2013 R1 (Lovejoy), 67P/Churyumov-Gerasimenko, and meteorites Murchinson and Murray \citep{co01,cro04, bi14, bi15, go15}.

In the ISM, HMCs are one of the earliest stages of the high-mass star-formation regions, and most of the complex and prebiotic molecules, including possible glycine (\ce{NH2CH2COOH}) precursors, are found in these objects \citep{van98, her09, gar13}. COMs in HMCs play a crucial role in increasing the chemical complexity in the ISM \citep{shi21}. HMCs are small, compact objects ($\leq$0.1 pc) with a warm temperature ($\geq$100 K) and a high gas density ($\geq$10$^{6}$ cm$^{-3}$) \citep{van98, wi14}. In the ISM, these objects are short-lived (10$^{5}$ yr to 10$^{6}$ yr) \citep{van98,vi04, gar06, gar08, gar13}. Most of the COMs are observed in the warm-inner regions of the envelope of HMCs, where the temperature exceeds the water ice desorption temperature of $\sim$100 K \citep{shi21}. The fractional abundances of the COMs in the HMCs are found to be between 10$^{-7}$ and 10$^{-11}$ with respect to molecular \ce{H2} \citep{her09}. In the HMCs, the gas temperature starts to increase due to the collapse of the pre-stellar core. This is called the "warm-up" phase \citep{vi04, gar13}. In this phase, several COMs are created in the ice, then sublimated and destroyed in the gas phase \citep{vi04}. So the resultant abundance of the complex molecules in the gas depends on the initial abundances of the precursors \citep{gar13}. Studies at the beginning of the 2000s showed that known (at the time) gas phase reactions were not efficient enough to explain the high abundances of COMs observed in HMCs. However, more recent quantum chemical studies have brought attention back to the gas phase chemistry in forming COMs \citep{ba15, sk18}.

The HMC candidate G358.93--0.03 MM1 is situated in the high-mass star-formation region G358.93--0.03 at a distance of 6.75$\,{}\,^{\,+0.37}_{\,-0.68}$ kpc \citep{re14,bro19}. The luminosity and total gas mass of G358.93--0.03 are $\sim$7.7$\times$10$^{3}$ \textup{L}$_{\odot}$ and 167$\pm$12 \textup{M}$_{\odot}$ \citep{bro19}. In this region, there exist a total of eight sub-millimeter continuum sources, which are defined as G358.93--0.03 MM1 to G358.93--0.03 MM8 in order of decreasing right ascension \citep{bro19, man23a, man23b}. Recently, \cite{man23b} found two more continuum sources associated with G358.93--0.03 MM1 and G358.93--0.03 MM2. Among the eight continuum sources, G358.93--0.03 MM1 and G358.93--0.03 MM3 are believed to be HMCs because the emission lines of \ce{CH3CN} and maser lines of \ce{CH3OH} are evident in those sources \citep{bro19, bay22}. The G358.93--0.03 MM1 hot core is more chemically rich than the G358.93--0.03 MM3 \citep{man23a, man23b}. Earlier, the maser lines of HDO, HNCO, and \ce{CH3OH} were detected towards the G358.93--0.03 MM1 using ALMA, TMRT, and VLA \citep{bro19, chen20}. Recently, the rotational emission lines of the possible urea precursor molecule cyanamide (\ce{NH2CN}), simplest sugar-like molecule glycolaldehyde (\ce{CH2OHCHO}), possible \ce{NH2CH2COOH} precursor molecule methylamine (\ce{CH3NH2}), peptide bond molecules formamide (\ce{NH2CHO}) and isocyanic acid (HNCO) were also detected from G358.93--0.03 MM1 using ALMA bands 6 and 7 \citep{man23a, man23b, man24b, man24c}.

In this article, we present the first detection of the simplest sugar alcohol molecule, EG, towards the HMC object G358.93--0.03 MM1 using the ALMA. In Section~\ref{obs}, we discuss the ALMA data and their reduction. The result of the identification and spatial distribution of EG is presented in Section~\ref{res}. The discussion and conclusions of the identification of EG are shown in Section~\ref{dis} and \ref{conclu}, respectively.

\begin{table*}
\centering
%\scriptsize 
\caption{Summary of the line parameters of the $aGg^{\prime}$-EG towards the G358.93--0.03 MM1.}
\begin{adjustbox} {width=1.0\textwidth}
\begin{tabular}{ccccccccccccccccc}
\hline 
Observed frequency &Transition & $E_{u}$ & $A_{ij}$ &$G_{up}$&FWHM &Optical depth&Peak intensity&Remark\\
				
(GHz) &(${\rm J_{K_a, K_c}}$, $v$) &(K)&(s$^{-1}$) &&(km s$^{-1}$) &($\tau$)&(K) & \\
\hline
290.543&28(6,22) $v$ = 0 -- 27(6,21) $v$ = 1&219.99&5.7$\times$10$^{-4}$&399&3.2$\pm$0.3&1.13$\times$10$^{-1}$&5.7$\pm$0.3&\bf{Non blended}\\
				
~~290.601$^{*}$&29(5,14) $v$ = 0 -- 28(15,13) $v$ = 1&324.24&4.39$\times$10$^{-4}$&531&3.2$\pm$0.3&7.04$\times$10$^{-2}$&6.4$\pm$0.2&\bf{Non blended}\\
				
290.614&21(3,18) $v$ = 0 -- 20(2,18) $v$ = 1&121.01&9.30$\times$10$^{-6}$&387&--&4.03$\times$10$^{-3}$&--&Blended with \ce{CH3OCHO}\\
				
~~290.772$^{*}$&29(14,15) $v$ = 0--28(14,14) $v$ = 1&310.05&4.61$\times$10$^{-4}$&531&3.2$\pm$0.9&6.47$\times$10$^{-2}$&5.7$\pm$0.6&\bf{Non blended}\\
				
290.811&44(15,29) $v$= 0 --44(14,30) $v$ = 0&597.41&6.64$\times$10$^{-5}$&623&--&1.74$\times$10$^{-3}$&--& Blended with \ce{C2H5CN}\\
				
290.811&44(15,30) $v$ = 0 -- 44(14,31) $v$ = 0&597.41&6.64$\times$10$^{-5}$&801&--&1.74$\times$10$^{-3}$&--&Blended with \ce{C2H5CN}\\
				
290.811&44(15,29) $v$ = 1 -- 44(14,30) $v$ = 1&597.67&7.02$\times$10$^{-5}$&801&--&1.74$\times$10$^{-3}$&--&Blended with \ce{C2H5CN}\\
				
290.812&44(15,30) $v$= 1 -- 44(14,31) $v$ = 1&597.67&7.03$\times$10$^{-5}$&623&--&1.74$\times$10$^{-3}$&--&Blended with \ce{C2H5CN}\\
				
290.826&29(5,25) $v$ =  0 -- 28(5,24) $v$ = 1&227.12&2.83$\times$10$^{-4}$&413&--&5.08$\times$10$^{-2}$&--& Blended with \ce{CH2(OH)CHO}\\
				
~~290.925$^{*}$&~32(1,32) $v$ = 0 -- 31(1,31) $v$ = 1&238.54&6.07$\times$10$^{-4}$&585&--&7.97$\times$10$^{-2}$&--&Blended with \ce{HCCCHO}\\
				
~~290.987$^{*}$&~~~29(13,16) $v$ = 0 -- 28(13,15) $v$ = 1&296.85&4.81$\times$10$^{-4}$&531&3.2$\pm$0.8&6.88$\times$10$^{-2}$&6.3$\pm$0.7&\bf{Non blended}\\
				
291.171&29(4,26) $v$ = 1 -- 28(3,25) $v$ = 1&220.92&1.18$\times$10$^{-4}$&531&--&2.75$\times$10$^{-2}$&--&Blended with \ce{CH2OHCHO}\\
				
291.221&29(4,26) $v$ = 0 -- 28(3,25) $v$ = 0&220.59&3.63$\times$10$^{-5}$&413&--&6.60$\times$10$^{-3}$&--&Blended with \ce{CH2OO}\\
				
291.263&29(12,18) $v$ = 0 -- 28(12,17) $v$ = 1&284.65&5.0$\times$10$^{-5}$&413&3.2$\pm$0.5&7.73$\times$10$^{-2}$&6.7$\pm$0.2& \bf{Non blended}\\
				
291.326&28(5,23) $v$ = 0 -- 27(5,22) $v$ = 1&216.39&5.94$\times$10$^{-5}$&399&3.2$\pm$0.5&1.39$\times$10$^{-1}$&7.0$\pm$0.2& \bf{Non blended}\\
				
291.390&43(15,28) $v$ = 0 -- 43(14,29) $v $ = 0&575.73&6.61$\times$10$^{-5}$&783&--&3.22$\times$10$^{-3}$&--&Blended with \ce{CH2OH}$^{13}$CHO\\
				
291.390&43(15,29) $v$ = 0 -- 43(14,30) $v$ = 0&575.73&6.61$\times$10$^{-5}$&609&--&3.22$\times$10$^{-3}$&--&Blended with \ce{CH2OH}$^{13}$CHO\\
				
291.395&43(15,28) $v$ = 1 -- 43(14,29) $v$ = 1&576.0&7.0$\times$10$^{-5}$&609&--&3.4$\times$10$^{-3}$&--&Blended with \ce{CH2OO}\\
				
291.395&43(15,29) $v$ = 1 -- 43(14, 30) $v$ = 1&576.0&7.0$\times$10$^{-5}$&783&--&3.4$\times$10$^{-3}$&--&Blended with \ce{CH2OO}\\
				
291.514&22(5,17) $v$ = 0 -- 21(4,17) $v$ = 1&138.24&1.32$\times$10$^{-5}$&315&--&5.57$\times$10$^{-3}$&--&Blended with \ce{DCOOH}\\
				
~~291.625$^{*}$&29(11,19) $v$ = 0 -- 28(11,18) $v$ = 1&273.46&5.19$\times$10$^{-5}$&413&3.2$\pm$0.6&6.68$\times$10$^{-2}$&9.0$\pm$0.3&\bf{Non blended}\\
				
291.628&27(6,21) $v$ = 1 -- 26(6,20) $v$ = 0&206.05&5.63$\times$10$^{-4}$&385&3.2$\pm$0.3&6.68$\times$10$^{-2}$&9.0$\pm$0.4& \bf{Non blended}\\
				
~~291.924$^{*}$&42(15,27) $v$ = 0 -- 42(14,28) $v$ = 0&554.55&6.58$\times$10$^{-5}$&595&--&3.58$\times$10$^{-3}$&--&Blended with \ce{CH3}$^{18}$OH\\
				
291.933&42(15,27) $v$ = 1 -- 42(14,28) $v$ = 1&554.82&6.98$\times$10$^{-5}$&765&--&4.87$\times$10$^{-3}$&--&Blended with H$^{13}$COOH\\
				
291.933&42(15,28) $v$ = 1 -- 42(14,20) $v$ = 1&554.82&6.98$\times$10$^{-5}$&595&--&4.87$\times$10$^{-3}$&--&Blended with H$^{13}$COOH\\

292.114&29(10,20) $v$ = 0 -- 28(10,19) $v$ = 1&263.30&5.64$\times$10$^{-4}$&413&--&1.49$\times$10$^{-1}$&--&Blended with \ce{CH3COOH}\\

292.115&29(10,19) $v$ = 0 -- 28(10,18) $v$ = 1&263.30&5.36$\times$10$^{-4}$&531&--&1.48$\times$10$^{-1}$&--&Blended with \ce{CH3COOH}\\

292.303&28(5,24) $v$ = 1 -- 27(5,23) $v$ = 0&213.16&6.37$\times$10$^{-4}$&399&3.2$\pm$0.7&2.82$\times$10$^{-1}$&6.2$\pm$0.5&\bf{Non blended}\\

292.627&30(2,28) $v$ = 0 -- 29(3,27) $v$ = 0&227.44&9.20$\times$10$^{-5}$&427&--&1.70$\times$10$^{-2}$&--&Blended with \ce{CH3OCHO}\\

292.633&30(3,27) $v$ = 0 -- 29(3,26) $v$ =1 &234.92&5.70$\times$10$^{-4}$&427&3.2$\pm$0.5&1.05$\times$10$^{-1}$&5.6$\pm$0.4&\bf{Non blended}\\

292.635&30(2,28) $v$ = 1 -- 29(3,27) $v$ = 1&227.77&8.91$\times$10$^{-5}$&549&3.2$\pm$0.5&1.03$\times$10$^{-1}$&5.6$\pm$0.4&\bf{Non blended}\\

292.699&30(3,28) $v$ = 0 -- 29(2,27) $v$ = 0&227.44&8.80$\times$10$^{-5}$&549&--&2.09$\times$10$^{-2}$&--&Blended with \ce{DCOOH}\\

292.704&30(3,28) $v$ = 1 -- 29(2,27) $v$ = 1 &227.77&9.33$\times$10$^{-5}$&427&--&2.08$\times$10$^{-2}$&--&Blended with \ce{CH3CHDCN}\\

292.794&29(9,21) $v$ = 0 -- 28(9,20) $v$ = 1&254.20&5.54$\times$10$^{-5}$&413&3.2$\pm$0.3&8.30$\times$10$^{-2}$&4.9$\pm$0.5&\bf{Non blended}\\

292.821&29(9,20) $v$ = 0 -- 28(9,19) $v$ = 1&254.20&5.54$\times$10$^{-4}$&531&3.2$\pm$0.5&1.06$\times$10$^{-1}$&4.8$\pm$0.6&\bf{Non blended}\\

293.177&30(2,29) $v$ = 1-- 29(2,28) $v$ = 0&219.56&6.15$\times$10$^{-4}$&427&3.2$\pm$0.9&1.59$\times$10$^{-1}$&6.1$\pm$0.2&\bf{Non blended}\\

293.178&30(1,29) $v$ = 1 -- 29(1,28) $v$ = 0&219.56&6.15$\times$10$^{-4}$&549&3.2$\pm$0.6&1.53$\times$10$^{-1}$&6.1$\pm$0.3&\bf{Non blended}\\

~~293.282$^{*}$&28(17,11) $v$ = 1 -- 27(17,10) $v$ = 0 &341.67&3.89$\times$10$^{-4}$&513&3.2$\pm$0.2&4.03$\times$10$^{-2}$&5.8$\pm$0.8&\bf{Non blended}\\

~~293.283$^{*}$&28(16,21) $v$ = 1 -- 27(16,11) $v$ = 0&325.49&4.15$\times$10$^{-4}$&513&3.2$\pm$0.2&4.02$\times$10$^{-2}$&5.8$\pm$0.8&\bf{Non blended}\\

293.299&39(15,24) $v$ = 1 -- 39(15,25) $v$ = 1&494.26&6.86$\times$10$^{-5}$&553&--&3.56$\times$10$^{-2}$&--&Blended with \ce{CH3OCHO}\\

~~293.308$^{*}$&28(18,10) $v$ = 1 -- 27(18,9) $v$ = 0 &358.82&3.62$\times$10$^{-4}$&513&--&3.34$\times$10$^{-2}$&--&Blended with \ce{CH3CH}$^{13}$CN\\

~~293.315$^{*}$&28(15,13) $v$ = 1 -- 27(15,12) $v$ = 0&310.30&4.40$\times$10$^{-4}$&513&--&5.61$\times$10$^{-2}$&--&Blended with \ce{E-CH3OH}\\

293.321&29(4,25) $v$ = 0 -- 28(4,24) $v$ = 1&226.66&6.45$\times$10$^{-4}$&531&3.2$\pm$0.9&1.48$\times$10$^{-1}$&5.2$\pm$0.3&\bf{Non blended}\\

~~293.356$^{*}$&28(19,9) $v$ = 1 -- 27(19,8) $v$ = 0&376.96&3.33$\times$10$^{-4}$&513&--&2.72$\times$10$^{-2}$&--&Blended with \ce{CH2}$^{13}$CHCN\\

~~293.386$^{*}$&28(14,14) $v$ = 1 -- 27(14,13) $v$ = 0&296.09&4.63$\times$10$^{-4}$&513&--&6.49$\times$10$^{-2}$&--&Blended with \ce{CH3OCHO}\\

~~293.424$^{*}$&28(20,8) $v$ = 1 -- 27(20,7) $v$ = 0&396.07&3.02$\times$10$^{-4}$&513&--&2.17$\times$10$^{-2}$&--&Blended with \ce{CH3CHDCN}\\

~~293.435$^{*}$&13(8,5) $v$ = 1 -- 12(7,5) $v$ = 0&76.49&2.01$\times$10$^{-5}$&189&--&5.77$\times$10$^{-3}$&--&Blended with \ce{C-C3H2}\\

~~293.506$^{*}$&27(13,16) $v$ = 1 -- 27(13,15) $v$ = 0&282.88&4.85$\times$10$^{-4}$&399&3.2$\pm$0.5&5.76$\times$10$^{-2}$&5.2$\pm$0.2&\bf{Non blended}\\

~~293.509$^{*}$&28(21,7) $v$ = 1 -- 27(21,6) $v$ = 0&416.16&2.70$\times$10$^{-4}$&513&3.2$\pm$0.5&5.70$\times$10$^{-2}$&5.2$\pm$0.2&\bf{Non blended}\\

~~293.574$^{*}$&11(9,2) $v$ = 1 -- 10(8,2) $v$ = 0&72.56&2.47$\times$10$^{-5}$&161&--&4.82$\times$10$^{-3}$&--&Blended with \ce{DNO3}\\

293.575&18(6,13) $v$ = 1 -- 17(5,12) $v$ = 1&102.29&3.90$\times$10$^{-5}$&259&--&4.80$\times$10$^{-3}$&--&Blended with \ce{DNO3}\\

293.608&26(5,22) $v$ = 0 -- 25(4,21) $v$ = 0&185.84&9.85$\times$10$^{-5}$&477&--&3.20$\times$10$^{-2}$&--&Blended with \ce{CH3OCHO}\\

~~293.610$^{*}$&28(22,6) $v$ = 1 -- 27(22,5) $v$ = 0&437.21&2.37$\times$10$^{-4}$&513&--&2.30$\times$10$^{-2}$&--&Blended with \ce{CH3CHO}\\

293.653&18(6,13) $v$ = 0 -- 17(5,12) $v$ = 0&101.96&5.46$\times$10$^{-5}$&333&--&1.80$\times$10$^{-2}$&--&Blended with \ce{C2H5CN}\\

293.658&38(15,23) $v$ = 0 -- 38(14,24) $v$ = 0&474.79&6.40$\times$10$^{-5}$&539&--&2.82$\times$10$^{-2}$&--&Blended with \ce{CH2CHDCN}\\

293.681&38(15,23) $v$ = 1 -- 38(14,24) $v$ = 1&475.07&6.81$\times$10$^{-5}$&693&--&3.52$\times$10$^{-2}$&--&Blended with \ce{CH2OHCHO}\\

293.692&28(12,17) $v$ = 1 -- 27(12,16) $v$ = 0&270.67&5.05$\times$10$^{-4}$&399&--&6.51$\times$10$^{-2}$&--&Blended with $^{13}$\ce{CH3CH2CN}\\

293.698&29(8,22) $v$ = 0 -- 28(8,21) $v$ = 1&246.18&5.70$\times$10$^{-4}$&413&--&8.95$\times$10$^{-2}$&--&Blended with \ce{CH2OHCHO}\\

293.725&28(23,5) $v$ = 1 -- 27(23,4) $v$ = 0&459.23&2.01$\times$10$^{-4}$&513&--&2.73$\times$10$^{-2}$&--&Blended with \ce{CH2}$^{13}$CHCN\\

~~293.853$^{*}$&28(24,4) $v$ = 1 -- 27(24,3) $v$ = 0&482.22&1.65$\times$10$^{-4}$&513&--&3.21$\times$10$^{-3}$&--&Blended with HC$^{34}$S\\

293.864&31(5,26) $v$ = 1 -- 30(6,25) $v$ = 1&262.32&8.08$\times$10$^{-5}$&441&--&8.05$\times$10$^{-2}$&--&Blended with \ce{CH3OCHO}\\

293.903&29(6,24) $v$ = 0 -- 28(6,23) $v$ = 1 &233.07&5.71$\times$10$^{-4}$&413&3.2$\pm$0.9&9.82$\times$10$^{-1}$&8.2$\pm$0.6&\bf{Non blended}\\

293.390&29(7,23) $v$ = 0 -- 28(7,22) $v$ = 1&239.22&4.32$\times$10$^{-4}$&413&3.2$\pm$0.6&2.01$\times$10$^{-1}$&5.5$\pm$0.3&\bf{Non blended}\\

293.966&28(11,18) $v$ = 1 -- 27(11,17) $v$ = 0 &259.46&5.25$\times$10$^{-4}$&399&3.2$\pm$0.3&3.52$\times$10$^{-1}$&5.6$\pm$0.5&\bf{Non blended}\\

294.012&29(8,21) $v$ = 0 -- 28(8,20) $v$ = 1 &246.21&5.72$\times$10$^{-4}$&531&3.2$\pm$0.4&8.20$\times$10$^{-1}$&5.1$\pm$0.2&\bf{Non blended}\\

294.195&27(5,22) $v$ = 1 -- 26(5,21) $v$ = 0&202.40& 5.93$\times$10$^{-4}$&385&--&2.58$\times$10$^{-3}$&--&Blended with \ce{C2H3CHO}\\

302.607&30(10,21) $v$ = 0 -- 29(10,2) $v$ = 1&278.13&6.02$\times$10$^{-4}$&549&3.2$\pm$0.9&1.62$\times$10$^{-1}$&8.9$\pm$0.9&\bf{Non blended}\\

302.610&30(10,20) $v$ = 0 -- 29(10,19) $v$ = 1&278.13&6.02$\times$10$^{-4}$&427&3.2$\pm$0.5&1.78$\times$10$^{-1}$&8.2$\pm$0.7&\bf{Non blended}\\

302.829&28(5,24) $v$ = 0 -- 27(4,23) $v$ = 0  &212.89&2.33$\times$10$^{-4}$&513&--&8.96$\times$10$^{-2}$&--&Blended with \ce{E-CH3OH}\\

302.903&30(7,24) $v$ = 0 -- 29(7,22) $v$ = 0 &254.16&1.56$\times$10$^{-4}$&549&--&4.92$\times$10$^{-2}$&--&Blended with \ce{CH3COOH}\\						
				%\hline
\end{tabular}	
\end{adjustbox}
\label{tab:MOLECULAR DATA}\\
		%*--The transition of $aGg^{\prime}$-EG contain double with frequency difference $\leq$100 kHz. The second transition is not shown.\\
\end{table*}

\begin{table*}
	%\begin{minipage}[t]{\columnwidth}
\centering
\scriptsize 
\text{{\large Table~1 Continued.}}
%\caption{Summary of the LTE fitted line parameters of the t-HC(O)SH towards the IRAS 16293 B.}
\begin{adjustbox}{width=1.0\textwidth}
\begin{tabular}{ccccccccccccccccc}
\hline 
Observed frequency &Transition & $E_{u}$ & $A_{ij}$ &$G_{up}$&FWHM &Optical depth&Peak intensity&Remark\\
			
(GHz) &(${\rm J_{K_a, K_c}}$, $v$) &(K)&(s$^{-1}$) &&(km s$^{-1}$) &($\tau$)&(K) & \\
\hline

303.240&28(6,22) $v$ = 1 -- 27(6,21) $v$ = 0&220.30&6.54$\times$10$^{-4}$&513&--&2.35$\times$10$^{-2}$&--&Blended with CH$_{3}$$^{13}$CH$_{2}$CN\\

303.361&30(9,22) $v$ = 0 -- 29(9,21) $v$ = 1 &269.06&6.21$\times$10$^{-4}$&543&--&1.77$\times$10$^{-1}$&--&Blended with \ce{HC5N}\\

303.409&30(9,21) $v$ = 0 -- 29(9,20) $v$ = 1&269.06&6.21$\times$10$^{-4}$&427&3.2$\pm$0.3&1.37$\times$10$^{-1}$&7.1$\pm$0.6& \bf{Non blended}\\

~~303.476$^{*}$& 29(17,12) $v$ = 1 -- 28(17,11)$v$ = 0&355.91&4.49$\times$10$^{-4}$&413&--&5.50$\times$10$^{-2}$&--&Blended with CH$_{2}$$^{13}$CHCN\\

~~303.488$^{*}$&29(16,13) $v$ = 1 -- 28(16,12) $v$ = 0&339.72&4.76$\times$10$^{-4}$&413&3.2$\pm$0.3&6.47$\times$10$^{-2}$&7.6$\pm$0.7&\bf{Non blended}\\

~~303.493$^{*}$&29(18,11) $v$ = 1 -- 28(18,10) $v$ = 0&373.03&4.20$\times$10$^{-4}$&413&--&4.61$\times$10$^{-2}$&--&Blended with \ce{CH3SH}\\

~~303.534$^{*}$&29(19,10) $v$ = 1 -- 28(19,9) $v$ = 0&391.21&3.90$\times$10$^{-4}$&413&3.2$\pm$0.6&7.51$\times$10$^{-2}$&9.5$\pm$0.8&\bf{Non blended}\\

~~303.596$^{*}$&29(20,9) $v$ = 1 -- 28(20,8) $v$ = 0&410.32&3.59$\times$10$^{-4}$&413&--&3.02$\times$10$^{-2}$&--&Blended with \ce{C2H3NC}\\

~~303.623$^{*}$&29(14,15) $v$ = 1 -- 28(14,14) $v$ = 0&410.35&5.25$\times$10$^{-4}$&413&3.2$\pm$0.5&8.62$\times$10$^{-2}$&7.8$\pm$0.5&\bf{Non blended}\\

~~303.676$^{*}$&29(21,8) $v$ = 0 -- 28(21,7) $v$ = 0&430.40&3.26$\times$10$^{-4}$&413&--&2.46$\times$10$^{-2}$&--&Blended with \ce{E-CH3OH}\\

303.766&29(13,17) $v$ = 1 -- 28(13,16) $v$ = 0&297.15&5.48$\times$10$^{-4}$&531&--&1.25$\times$10$^{-1}$&--&Blended with \ce{CH3OCHO}\\

303.773&29(22,7) $v$ = 1 -- 28(22,6) $v$ = 0&451.46&2.91$\times$10$^{-4}$&413&--&1.92$\times$10$^{-2}$&--&Blended with \ce{CH3CHDCN}\\

303.849&10(10,0) $v$ = 1 -- 9(9,0) $v$ = 0 &76.50&3.27$\times$10$^{-4}$&189&--&1.11$\times$10$^{-2}$&--&Blended with \ce{CH2OHCHO}\\

~~303.886$^{*}$&29(23,6) $v$ = 1 -- 28(23,5) $v$ = 0&473.48&2.55$\times$10$^{-4}$&413&--&1.45$\times$10$^{-2}$&--&Blended with \ce{CH3OCHO}\\

303.918&30(6,25) $v$ = 0 -- 29(6,24) $v$ = 1&247.94&6.25$\times$10$^{-4}$&549&3.2$\pm$0.7&2.04$\times$10$^{-1}$&8.8$\pm$0.2&\bf{Non blended}\\

~~303.981$^{*}$&29(12,18) $v$ = 1 -- 28(12,17) $v$ = 0&284.95&5.69$\times$10$^{-4}$&531&3.2$\pm$0.6&1.10$\times$10$^{-1}$&9.5$\pm$0.6&\bf{Non blended}\\

~~304.012$^{*}$&29(24,5) $v$ = 1 -- 28(24,4) $v$ = 0 &496.47&2.17$\times$10$^{-4}$&413&--&1.06$\times$10$^{-2}$&--&Blended with \ce{C2H5}C$^{15}$N\\

304.062&32(3,30) $v$ = 0 -- 31(3,29) $v$ = 1&256.85&6.82$\times$10$^{-4}$&585&3.2$\pm$0.9&2.23$\times$10$^{-1}$&9.2$\pm$0.6&\bf{Non blended}\\

304.062&32(2,30) $v$ = 0 -- 31(2,29) $v$ = 1&256.85&6.80$\times$10$^{-4}$&455&--&4.07$\times$10$^{-2}$&--&Blended with \ce{CH3CHO}\\

~~304.152$^{*}$&29(25,4) $v$ = 1 -- 28(25,3) $v$ = 0&520.42&1.77$\times$10$^{-4}$&413&--&7.43$\times$10$^{-3}$&--&Blended with \ce{C2H3CN}\\

~~304.296$^{*}$&29(11,19) $v$ = 1 -- 28(11,18) $v$ = 0&273.76&5.90$\times$10$^{-4}$&531&3.2$\pm$0.7&1.22$\times$10$^{-1}$&8.5$\pm$0.9&\bf{Non blended}\\

304.308&28(5,23) $v$ = 1 -- 27(5,22) $v$ = 0 &216.70&6.53$\times$10$^{-4}$&513&--&2.43$\times$10$^{-1}$&--&Blended with \ce{H2CS}\\

304.317&30(8,23) $v$ = 0 -- 29(8,22) $v$ = 1&261.08&6.35$\times$10$^{-4}$&549&--&1.89$\times$10$^{-1}$&--&Blended with \ce{C2H5OCHO}\\

304.482&29(4,25) $v$ = 1 -- 28(4,24) $v$ = 0&226.91&3.28$\times$10$^{-4}$&413&--&9.19$\times$10$^{-2}$&--&Blended with \ce{CH2}$^{13}$CHCN\\

304.515&32(1,31) $v$ = 0 -- 31(2,30) $v$ = 0&248.01&1.22$\times$10$^{-4}$&455&--&4.20$\times$10$^{-2}$&--&Blended with \ce{c-C2H4O}\\

304.516&32(2,31) $v$ = 0 -- 31(1,30) $v$ = 0&248.01&1.21$\times$10$^{-4}$&585&--&4.22$\times$10$^{-2}$&--&Blended with \ce{c-C2H4O}\\

304.526&32(1,31) $v$ = 1 -- 31(2,30) $v$ = 0&248.35&1.22$\times$10$^{-4}$&585&--&4.28$\times$10$^{-2}$&--&Blended with \ce{C2H5}$^{13}$CN\\

304.526&32(2,31) $v$ = 1 -- 31(1,30) $v$ = 1&248.35&1.22$\times$10$^{-4}$&455&--&3.27$\times$10$^{-2}$&--&Blended with \ce{C2H5}$^{13}$CN\\

304.754&29(10,20) $v$ = 1 -- 28(10,19) $v$ = 0&263.60&6.10$\times$10$^{-4}$&531&--&8.20$\times$10$^{-2}$&--&Blended with \ce{CH3COCH3}\\

304.756&29(10,19) $v$ = 1 -- 28(10,18) $v$ = 1&263.60&6.10$\times$10$^{-4}$&413&--&6.32$\times$10$^{-2}$&--&Blended with \ce{CH3COCH3}\\

304.811&30(8,22) $v$ = 0 -- 29(8,21) $v$ = 1 &261.15&6.38$\times$10$^{-4}$&427&--&2.38$\times$10$^{-2}$&--&Blended with \ce{C2H3CHO}\\

305.235&32(1,32) $v$ = 1 -- 31(1,31) $v$ = 0&238.88&7.02$\times$10$^{-4}$&455&3.2$\pm$0.9&1.99$\times$10$^{-1}$&6.5$\pm$0.2&\bf{Non blended}\\

305.423&29(6,21) $v$ = 1 -- 28(9,20) $v$ = 0&254.50&6.30$\times$10$^{-4}$&531&--&1.88$\times$10$^{-1}$&--&Blended with \ce{C-C3H2}\\

305.454&29(9,20) $v$ = 1 -- 28(9,19) $v$ = 0&254.50&6.30$\times$10$^{-4}$&413&--&2.38$\times$10$^{-1}$&--&Blended with \ce{SO2}\\

305.729&29(6,24) $v$ = 1 -- 28(6,23) $v$ = 0&233.35&6.80$\times$10$^{-4}$&531&--&2.33$\times$10$^{-2}$&--&Blended with \ce{CH3OCHO}\\

305.804&30(7,24) $v$ = 0 -- 29(7,23) $v$ = 1&254.16&4.95$\times$10$^{-4}$&549&--&1.53$\times$10$^{-1}$&--&Blended with \ce{C2H3CN}\\
\hline
\end{tabular}	
\end{adjustbox}
*--The transition of $aGg^{\prime}$-EG contain double with frequency difference $\leq$100 kHz. The second transition is not shown.\\
%		\end{minipage}[t]{\columnwidth}
\end{table*}

\section{Observations and data reduction}
\label{obs}
The high-mass star-formation region G358.93--0.03 was observed using ALMA band 7 receivers to study the massive protostellar accretion outburst (ID: 2019.1.00768.S., PI: Crystal Brogan). The observations were performed on October 11th, 2019 with on-source integration times of 756 s. The observed phase center of G358.93--0.03 was ($\alpha,\delta$)$_{\rm J2000}$ = 17:43:10.000, --29:51:46.000. A total of 47 antennas were used to observe the G358.93--0.03 with a minimum baseline of 14 m and a maximum baseline of 2517 m. During the observation, J1550+0527 was used as a flux calibrator and bandpass calibrator, and J1744--3116 was used as a phase calibrator. The observed frequency ranges of G358.93--0.03 were 290.51--292.39 GHz, 292.49--294.37 GHz, 302.62--304.49 GHz, and 304.14--306.01 GHz, with a spectral resolution of 977 kHz.

For data reduction and imaging, we used the Common Astronomy Software Application ({\tt CASA 5.4.1}) with the ALMA data analysis pipeline \citep{mc07}. In the pipeline, we used the task {\tt SETJY} with the Perley-Butler 2017 flux calibrator model for flux calibration \citep{pal17}. We also applied the pipeline tasks {\tt hifa\_bandpassflag} and {\tt hifa\_flagdata} for bandpass calibration and flagging the bad antenna data. After the initial data reduction, we used the task {\tt MSTRANSFORM} to separate the target object G358.93--0.03. We created the four continuum images of G358.93--0.03 by using line-free channels at frequency ranges of 290.51--292.39 GHz, 292.49--294.37 GHz, 302.62--304.49 GHz, and 304.14--306.01 GHz using the CASA task {\tt TCLEAN} with the {\tt HOGBOM} deconvolver. Recently, \citet{man23b} analyzed this data, and they showed the continuum emission image of G358.93--0.03 at frequency 303.39 GHz (988 $\mu$m) (Fig. 1. of \citet{man23b}). \citet{man23b} clearly detected the eight sub-millimeter wavelength continuum sources, G358.93--0.03 MM1 to G358.93--0.03 MM8. \citet{man23b} also detected other two continuum sources associated with G358.93--0.03 MM1 and G358.93--0.03 MM2, which are defined as G358.93--0.03 MM1A and G358.93--0.03 MM2A. Before making the spectral images, we use the task {\tt UVCONTSUB} to subtract the background continuum emission from the UV plane of the calibrated data. We also made self-calibration multiple times using the tasks {\tt GAINCAL} and {\tt APPLYCAL} for better RMS of the final images. We create the spectral images of G358.93--0.03 at frequency ranges of 290.51--292.39 GHz, 292.49--294.37 GHz, 302.62--304.49 GHz, and 304.14--306.01 GHz using the task {\tt TCLEAN} with the {\tt SPECMODE} = {\tt CUBE} parameter. Finally, we apply the task {\tt IMPBCOR} for the correction of the primary beam pattern in the spectral images of G358.93--0.03.

\section{Results}
\label{res}

\subsection{Line emission from the G358.93--0.03}
In the spectral images of G358.93--0.03, we observe that only the spectra of G358.93--0.03 MM1 and G358.93--0.03 MM3 show any line emission. The synthesized beam sizes of the spectral images of star-forming region G358.93--0.03 at frequency ranges of 290.51--292.39 GHz, 292.49--294.37 GHz, 302.62--304.49 GHz, and 304.14--306.01 GHz are 0.42 arcsec $\times$ 0.36 arcsec, 0.42 arcsec $\times$ 0.37 arcsec, 0.41 arcsec $\times$ 0.36 arcsec, and 0.41 arcsec $\times$ 0.35 arcsec, respectively. We generate the molecular spectra from the HMCs G358.93--0.03 MM1 and G358.93--0.03 MM3 by creating a 0.91 arcsec diameter circular region, which is larger than the line-emitting regions of these two sources. The extracted molecular spectra of G358.93--0.03 MM1 and G358.93--0.03 MM3 in the observable frequency ranges are shown in Fig. 2 of \citet{man23b}. We notice that the molecular spectra of G358.93--0.03 MM1 are more chemically rich than the spectra of G358.93--0.03 MM3. In the molecular spectra of G358.93--0.03 MM1, we notice the evidence of an inverse P Cygni profile associated with the \ce{CH3OH} emission lines. The presence of the inverse P Cygni profile indicates that the G358.93--0.03 MM1 is undergoing infall. The position of G358.93--0.03 MM1 is RA (J2000) = 17$^{h}$43$^{m}$10$^{s}$.101, Dec (J2000) = --29$^\circ$51$^{\prime}$45$^{\prime\prime}$.693. The position of G358.93--0.03 MM3 is RA (J2000) = 17$^{h}$43$^{m}$10$^{s}$.0144, Dec (J2000) = --29$^\circ$51$^{\prime}$46$^{\prime\prime}$.193. The systemic velocities ($V_{LSR}$) of the spectra of G358.93--0.03 MM1 and G358.93--0.03 MM3 are --16.5 km s$^{-1}$ and --18.2 km s$^{-1}$, respectively \citep{bro19}.

\begin{figure*}
\centering
\includegraphics[width=0.93\textwidth]{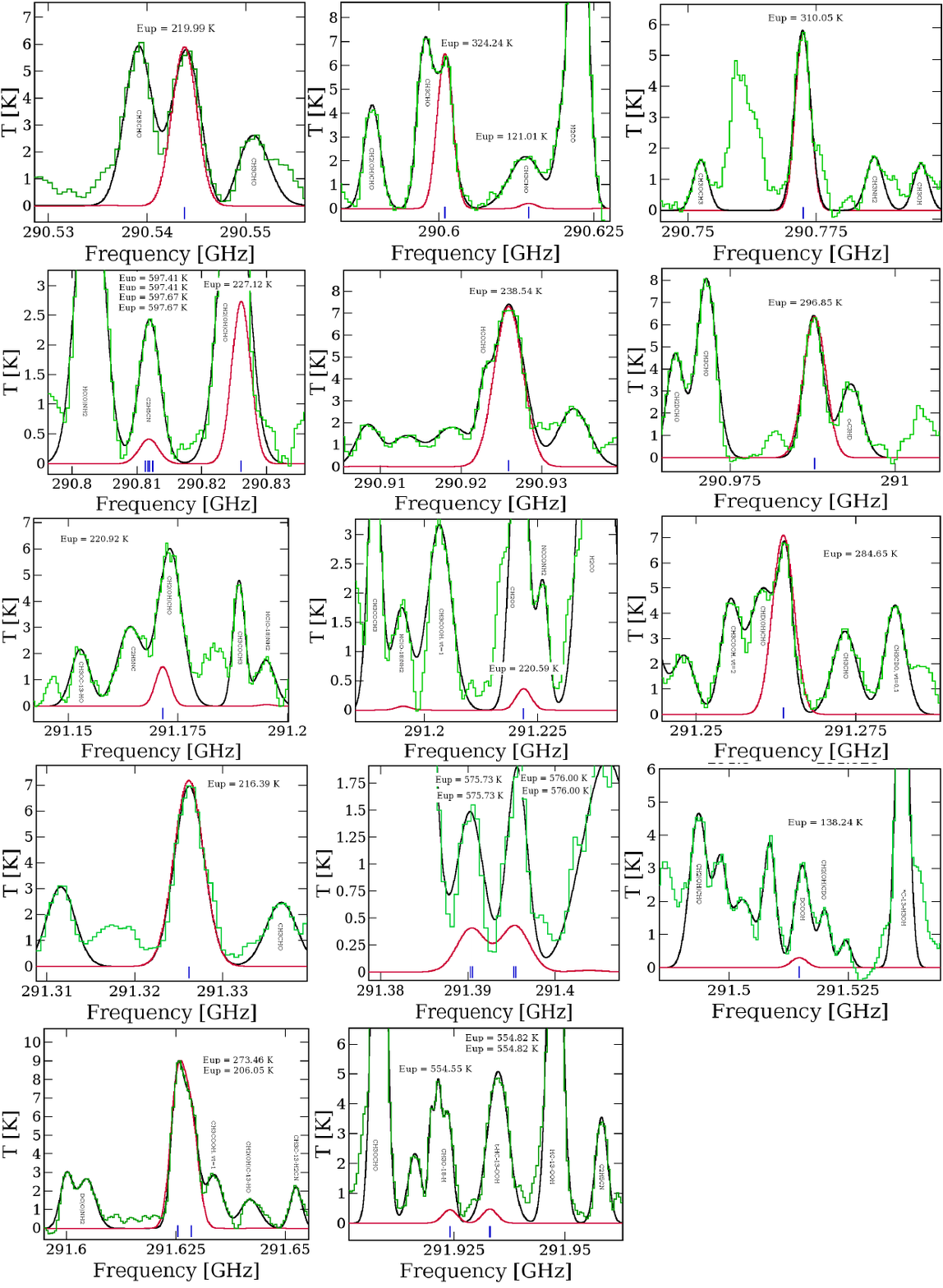}
\caption{Rotational emission lines of $aGg^{\prime}$-EG towards the G358.93--0.03 MM1. The green spectra indicate the observed molecular spectra of the G358.93--0.03 MM1. The red spectra present the LTE model spectrum of $aGg^{\prime}$-EG and the black spectra indicate the model spectra for all species identified in the spectrum, including those published in \citet{man23b}, \citet{man24b}, and \citet{man24c}.}
\label{fig:ltespec}
\end{figure*}
\begin{figure*}
\text{{\large Figure~2 Continued.}}
\centering
\includegraphics[width=0.93\textwidth]{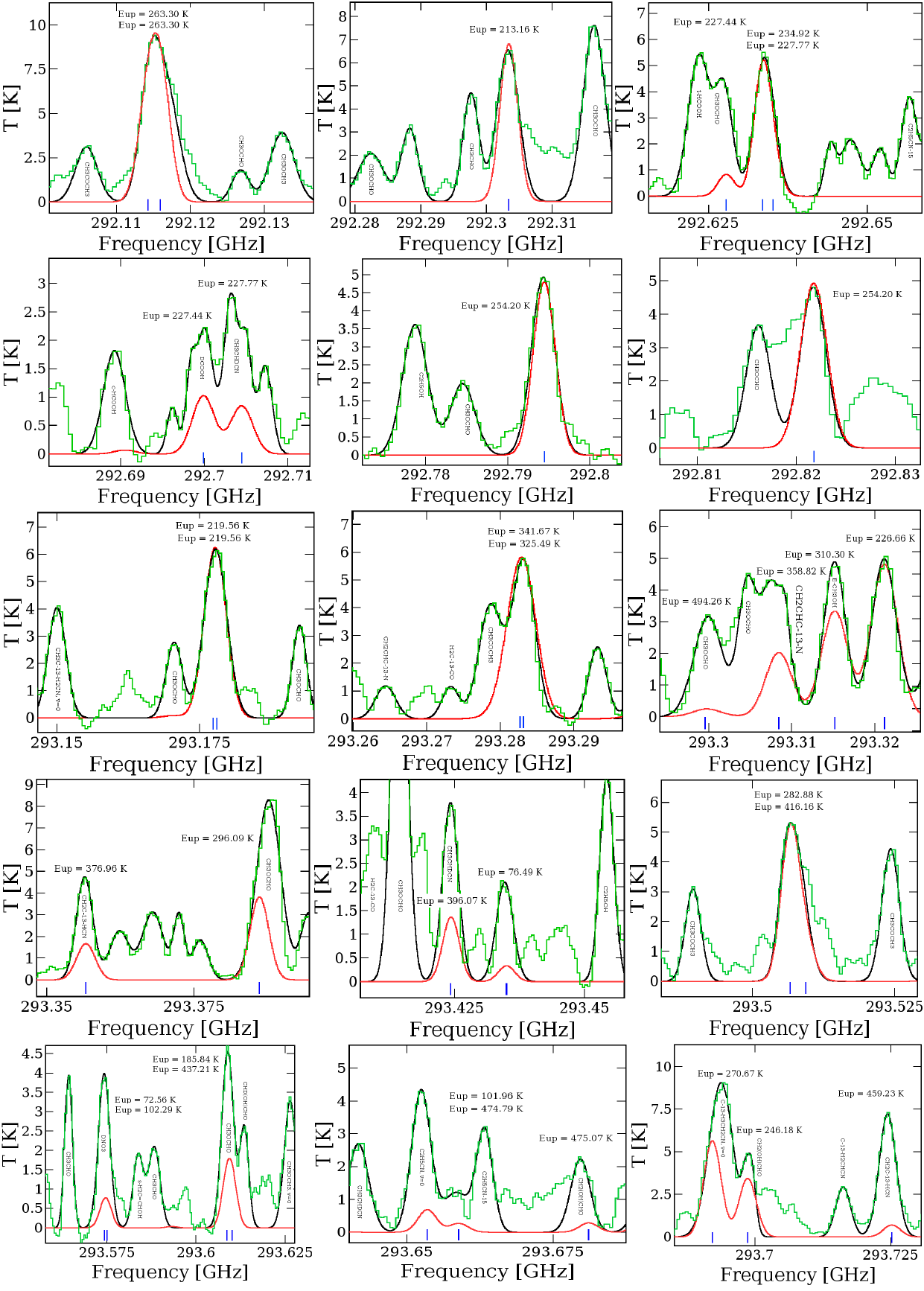}
\end{figure*}
\begin{figure*}
\text{{\large Figure~2 Continued.}}
\centering
\includegraphics[width=0.93\textwidth]{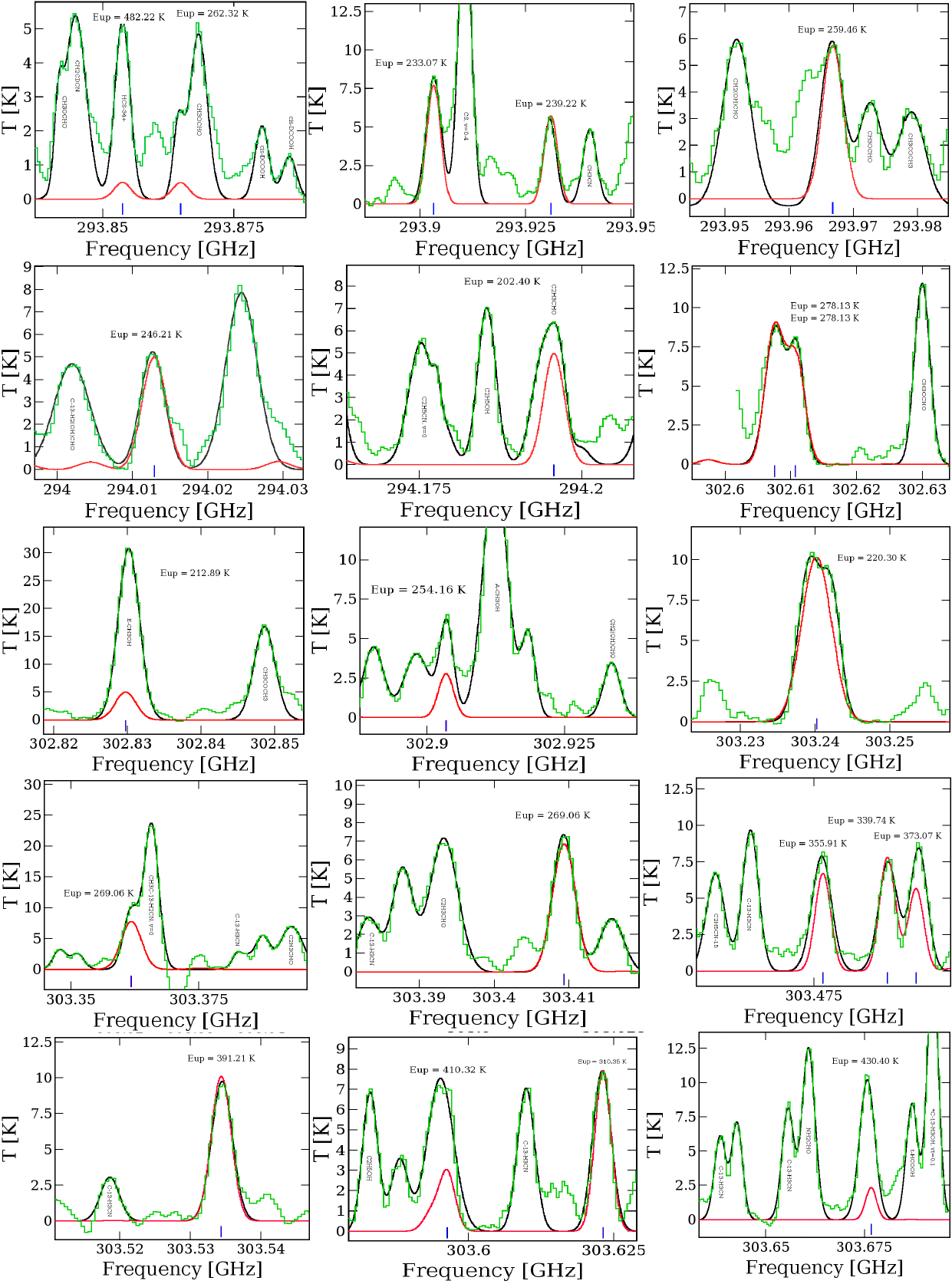}
\end{figure*}
\begin{figure*}
\text{{\large Figure~2 Continued.}}
\centering
\includegraphics[width=0.93\textwidth]{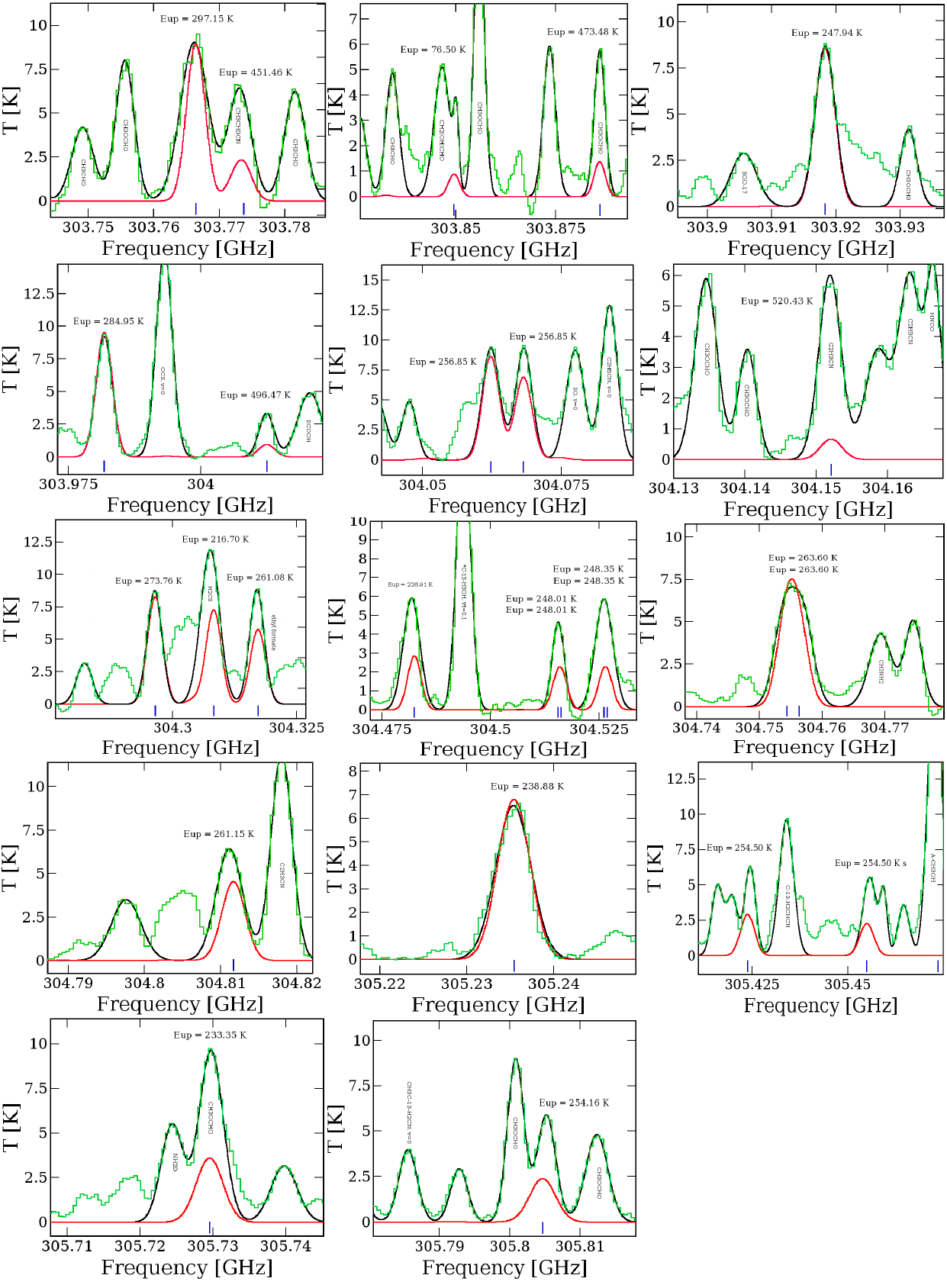}
\end{figure*}

\begin{table}
\centering
\scriptsize 
\caption{Emitting regions of $aGg^{\prime}$-EG towards the G358.93--0.03 MM1.}
\begin{adjustbox}{width=0.48\textwidth}
\begin{tabular}{ccccccccccccccccc}
\hline 
Observed frequency&Transition&$E_{u}$ &Emitting region\\
(GHz)            &(${\rm J_{K_a, K_c}}$, $v$)          & (K)     &(arcsec) \\
\hline
290.543&28(6,22) $v$ = 0 -- 27(6,21) $v$ = 1&219.99&0.41\\
~~290.772$^{*}$&29(14,15) $v$ = 0 -- 28(14,14) $v$ = 1&310.05&0.41\\
291.263&29(12,18) $v$ = 0 -- 28(12,17) $v$ = 1&284.65&0.42\\
291.326&28(5,23) $v$ = 0 -- 27(5,22) $v$ = 1&216.39&0.41\\
292.821&29(9,20) $v$ = 0 -- 28(9,19) $v$ = 1&254.20&0.41\\
293.903&29(6,24) $v$ = 0 -- 28(6,23) $v$ = 1&233.07&0.41\\
293.966&28(11,18) $v$ = 1 -- 27(11,17) $v$ = 0&259.46&0.42\\
303.409&30(9,21) $v$ = 0 -- 29(9,20) $v$ = 1&269.06&0.41\\
303.918&30(6,25) $v$ = 0 -- 29(6,24) $v$ = 1&247.94&0.40\\
~~303.981$^{*}$&29(12,18) $v$ = 1 -- 28(12,17) $v$ = 0&284.95&0.40\\
304.062&32(3,30) $v$ = 0 -- 31(3,29) $v$ = 1&256.85&0.41\\
305.235&32(1,32) $v$ = 1 -- 31(1,31) $v$ = 0&238.88&0.42\\
\hline
\end{tabular}	
\end{adjustbox}
\label{tab:emitting region}
*--The transition of $aGg^{\prime}$-EG contain double with frequency difference $\leq$100 kHz. The second transition is not shown.\\
\end{table}

\subsubsection{Detection of EG towards the G358.93--0.03 MM1}
\label{sec:fitting}
We assume local thermodynamic equilibrium (LTE) and used the Cologne Database for Molecular Spectroscopy (CDMS) \citep{mu05} to identify the rotational emission lines of EG from the molecular spectra of G358.93--0.03 MM1. We use CASSIS for the LTE modelling \citep{vas15}. The LTE assumption is reasonable in the inner region of the G358.93--0.03 MM1 because the maximum gas density of the warm inner region of the hot core is 1$\times$10$^{8}$ cm$^{-3}$ \citep{ste21}. We used the Markov Chain Monte Carlo (MCMC) algorithm in CASSIS for fitting the LTE model spectra of EG over the observed molecular spectra of G358.93--0.03 MM1. The details of the MCMC fitting method using the CASSIS are described in \citet{gor20}. After the LTE analysis, we have detected a total of 106 transitions of the $aGg^{\prime}$ conformer of EG with $\geq$2.5$\sigma$ statistical significance in the spectra of G358.93--0.03 MM1 in the observable frequency ranges. After the spectral analysis using the LTE model, we observe that 35 transitions of $aGg^{\prime}$-EG are non-blended, and all of those non-blended transitions are detected above 5$\sigma$. To understand the non-blended transitions of $aGg^{\prime}$, we have used more than 200 different molecular transitions in the LTE modelling, which are taken from the CDMS database, including those molecules detected by \citet{man23a}, \citet{man23b}, \citet{man24b}, and \citet{man24c}. We find that the 71 transition lines of $aGg^{\prime}$-EG are blended with the other nearby molecular transitions. The upper-level energies of the detected 106 transitions of $aGg^{\prime}$-EG vary from 76.49 K to 597.41 K. The upper-level energies of the non-blended transitions of $aGg^{\prime}$-EG vary between 206.05 K and 416.16 K. There are no missing high-intensity transitions of $aGg^{\prime}$-EG in the observable frequency ranges. After the LTE modelling, the best-fit column density of $aGg^{\prime}$-EG is (4.5$\pm$0.1)$\times$10$^{16}$ cm$^{-2}$ with an excitation temperature of 155$\pm$35 K and a source size of 0.50 arcsec. During the LTE fitting, we used the FWHM of the LTE spectra of $aGg^{\prime}$-EG is 3.21 km s$^{-1}$. We have taken this FWHM value for LTE modelling of $aGg^{\prime}$-EG because recently \citet{man23b} estimated the FWHM of \ce{CH2OHCHO} in the spectra of G358.93--0.03 MM1 using this data was 3.2 km s$^{-1}$. The LTE-fitted rotational emission spectra of $aGg^{\prime}$-EG towards the G358.93--0.03 MM1 are shown in Fig.~\ref{fig:ltespec}. The LTE-fitted spectral line parameters of $aGg^{\prime}$-EG are shown in Tab.~\ref{tab:MOLECULAR DATA}. After the detection of the emission lines $aGg^{\prime}$-EG from the spectra of G358.93--0.03 MM1, we also search the emission lines of $gGg^{\prime}$-EG using the LTE modelling spectra. After the spectral analysis, we observe that all detected lines of $gGg^{\prime}$-EG are blended with other nearby molecular transitions. Thus, all detected lines of $gGg^{\prime}$-EG are blended with other molecules, so we estimate the upper limit column density of $gGg^{\prime}$-EG towards the G358.93--0.03 MM1 is $\leq$(2.9$\pm$0.6)$\times$10$^{16}$ cm$^{-2}$. \citet{man23b} estimated that the temperature of GA is 300 K, and we estimate that the temperature of EG towards the G358.93--0.03 MM1 is 155 K. Previously, \citet{shi21} showed that the temperatures of complex molecules vary between 100 K to 300 K in the inner shell of hot molecular core (see Fig 8 in \citet{shi21}). So, both molecules EG and GA exist in slightly different regions of the inner shell of G358.93--0.03 MM1.

\subsubsection{Abundance of EG towards the G358.93--0.03 MM1}
To determine the fractional abundance of $aGg^{\prime}$-EG, we use the column density of $aGg^{\prime}$-EG inside the 0.50 arcsec beam, which we divide by the molecular \ce{H2} column density. The fractional abundance of $aGg^{\prime}$-EG with respect to molecular \ce{H2} towards the G358.93--0.03 MM1 is (1.4$\pm$0.5)$\times$10$^{-8}$, where the column density of \ce{H2} towards the G358.93--0.03 MM1 is (3.1$\pm$0.2)$\times$10$^{24}$ cm$^{-2}$ which is derived from the dust continuum emission at wavelength 988 $\mu$m \citep{man23b}. The column density ratio of $aGg^{\prime}$-EG and GA (hereafter EG/GA) is 3.1$\pm$0.5, where the column density of GA towards the G358.93--0.03 MM1 is (1.5$\pm$0.9)$\times$10$^{16}$ cm$^{-2}$ \citep{man23b}. After spectral analysis using the LTE model, we also find the column density and excitation temperature of \ce{CH3OH} towards G358.93--0.03 MM1, which are (7.5$\pm$0.5)$\times$10$^{18}$ cm$^{-2}$ and 150$\pm$28 K, respectively. So, the column density ratio of $aGg^{\prime}$-EG and \ce{CH3OH} is (6.1$\pm$0.3)$\times$10$^{-3}$. \citet{mondal21} and \citet{min23} derived the abundance of $aGg^{\prime}$-EG with respect to \ce{H2} towards two other hot molecular cores, G10.47+0.03 and G31.41+0.31, to be (3.7$\pm$1.5)$\times$10$^{-8}$ and (1.5$\pm$0.4)$\times$10$^{-8}$ respectively, which are close to our derived abundance of $aGg^{\prime}$-EG towards the G358.93--0.03 MM1. This may indicate that the chemical formation pathway(s) of $aGg^{\prime}$-EG towards the G358.93--0.03 MM1 may be similar to those in G10.47+0.03 and G31.41+0.31.

\subsubsection{Searching for EG towards the G358.93--0.03 MM3}
We also attempt to search for emission lines of $aGg^{\prime}$-EG and $gGg^{\prime}$-EG towards the G358.93--0.03 MM3 using the LTE-modelled spectra, but we do not detect them. The derived upper limit column densities of $aGg^{\prime}$-EG and $gGg^{\prime}$-EG towards the G358.93--0.03 MM3 are $\leq$(1.6$\pm$0.9)$\times$10$^{15}$ cm$^{-2}$ and $\leq$(1.2$\pm$0.5)$\times$10$^{15}$ cm$^{-2}$ respectively. We assumed the excitation temperature of 155 K to estimate the upper limit column density of EG towards the G358.93--0.03 MM3. The upper limits of their fractional abundance are $\leq$(4.5$\pm$1.2)$\times$10$^{-9}$ and $\leq$(3.4$\pm$0.7)$\times$10$^{-9}$ respectively, where the column density of \ce{H2} towards the G358.93--0.03 MM3 is (3.5$\pm$0.7)$\times$10$^{23}$ cm$^{-2}$ \citep{man23b}.

\begin{figure*}
\centering
\includegraphics[width=1.0\textwidth]{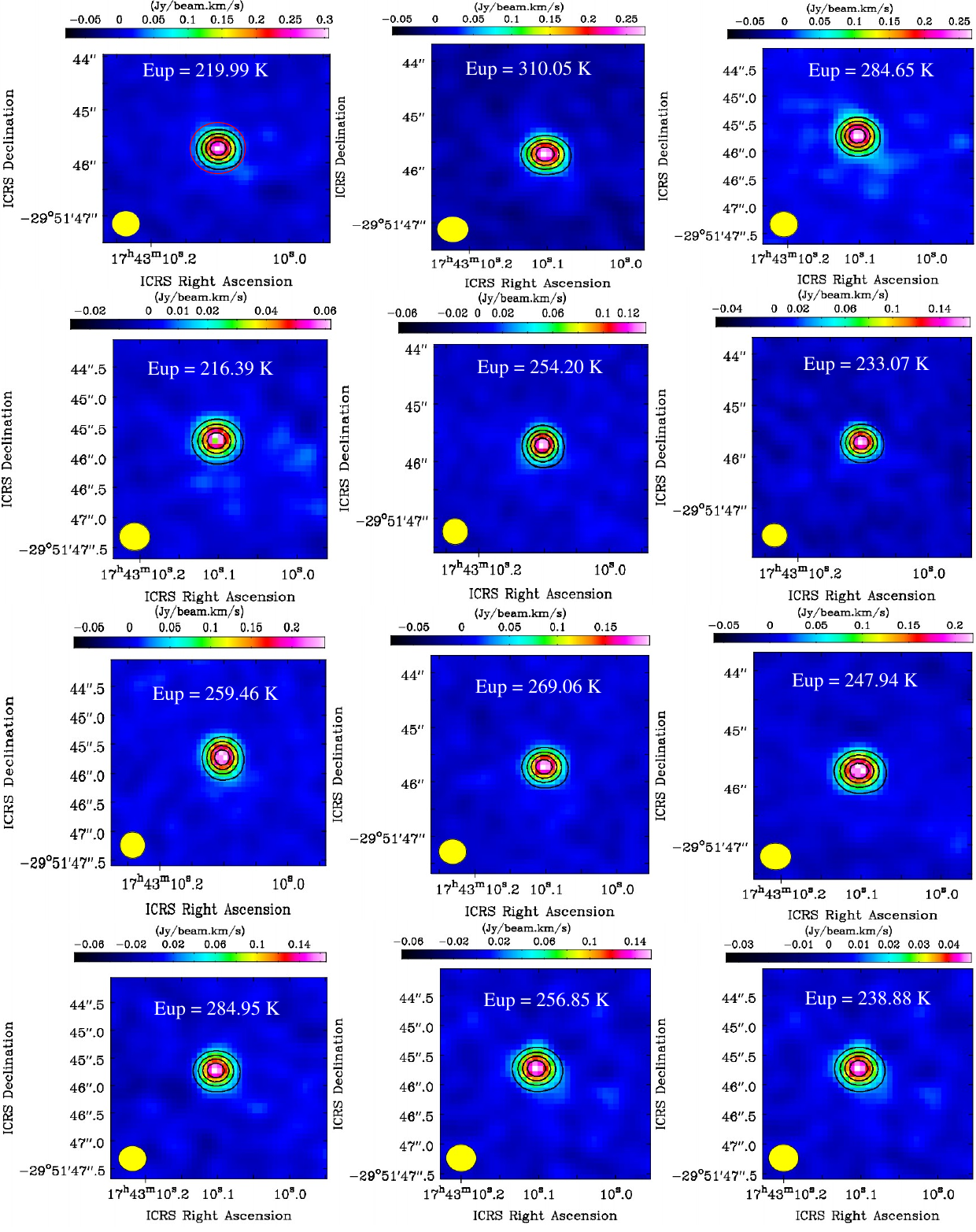}
\caption{Integrated emission maps (moment zero) of $aGg^{\prime}$-EG towards the G358.93--0.03 MM1. The integrated emission maps are overlaid with the 988 $\mu$m continuum emission map of G358.93--0.03 (black contour). The contour levels are at 20\%, 40\%, 60\%, and 80\% of the peak flux. The yellow circles represent the synthesized beams of the integrated emission maps. In the first panel image, the red circle indicates the 0.91 arcsec diameter circular region.}
\label{fig:emissionmap}
\end{figure*}

\subsection{Spatial distribution of $aGg^{\prime}$-EG towards the G358.93--0.03 MM1}
We produce the integrated emission maps (moment zero maps) of some selected highly intense non-blended emission lines of $aGg^{\prime}$-EG towards the G358.93--0.03 MM1 using the task {\tt IMMOMENTS}. During the run of the task {\tt IMMOMENTS}, we use the channel ranges of the spectral images where the emission lines of $aGg^{\prime}$-EG are detected. We create the integrated emission maps of $aGg^{\prime}$-EG for non-blended transitions towards the G358.9--0.03 MM1 at Fig.~\ref{fig:emissionmap}. We overlaid the 988 $\mu$m continuum emission map of G358.93--0.03 over the integrated emission maps of $aGg^{\prime}$-EG. The continuum emission map of G358.93--0.03 is taken from \citet{man23b}. We found that the integrated emission maps of $aGg^{\prime}$-EG exhibit a peak at the position of the continuum. The integrated emission maps indicate that the emission lines of $aGg^{\prime}$-EG originate from the high density, the warm inner region of G358.93--0.03 MM1. After the extraction, we apply the CASA task IMFIT to fit the 2D Gaussian over the integrated emission maps of $aGg^{\prime}$-EG to estimate the size of the emitting regions. The following equation is used
	
\begin{equation}
\theta_{S}=\sqrt{\theta^2_{50}-\theta^2_{\text{beam}}}
\end{equation}
where $\theta_{\text{beam}}$ indicates the half-power width of the synthesised beam and $\theta_{50} = 2\sqrt{A/\pi}$ denotes the diameter of the circle whose area surrounded the $50\%$ line peak of $aGg^{\prime}$-EG \citep{riv17}. The derived emitting regions of $aGg^{\prime}$ at several frequencies are shown in Tab.~\ref{tab:emitting region}. The synthesized beam sizes of the integrated emission maps vary between 0.41 arcsec $\times$ 0.35 arcsec and 0.42 arcsec $\times$ 0.36 arcsec. The derived emitting regions of $aGg^{\prime}$-EG at different frequencies vary between 0.40 arcsec and 0.42 arcsec. After fitting a 2D Gaussian, we observe that the size of the emitting regions of $aGg^{\prime}$-EG is comparable to or slightly greater than the synthesized beam sizes of the integrated emission maps. This result indicates that the detected transition lines of $aGg^{\prime}$-EG are not spatially resolved or are only marginally resolved towards the G358.93--0.03 MM1. As a result, drawing any conclusions about the morphology of the spatial distributions of $aGg^{\prime}$-EG towards the G358.93--0.03 MM1 is impossible. Higher spatial and angular resolution observations are required to understand the spatial distribution of $aGg^{\prime}$-EG towards the G358.93--0.03 MM1.

\begin{table}
\centering
\caption{Summary of the EG/GA ratios in the star-forming region.}
\begin{adjustbox}{width=0.48\textwidth}
\begin{tabular}{ccccccccccccccccc}
\hline 
Source&Luminosity&EG/GA&Referance\\
	  &(\textup{L}$_{\odot}$)& & \\
\hline
IRAS16293 B&$<$21&1.6$\pm$0.3&\cite{jo16}	\\
NGC1333 IRAS2A&~20&5$\pm$1&\cite{cou15}\\
NGC7129 FIRS2&~500&2.1$\pm$0.5&\cite{fu14}\\
G358.93--0.03 MM1&7.7$\times$10$^{3}$&3.1$\pm$0.5&This work\\
Orion KL&1.0$\times$10$^{5}$&$>$13&\cite{bro15}\\
G31.41+0.31&1.8$\times$10$^{5}$&10$\pm$1&\cite{riv17}\\
G34.3+0.2&2.8$\times$10$^{5}$&$>$6.3&\cite{ly15}\\
W51e2&4.7$\times$10$^{6}$&$>$15.5&\cite{ly15}\\
\hline
\end{tabular}	
\end{adjustbox}
\label{tab:ratio}
\end{table}

\section{Discussion}
\label{dis}
\subsection{Comparison of the EG/GA ratio in different star-forming regions}
\begin{figure}
\centering
\includegraphics[width=0.46\textwidth]{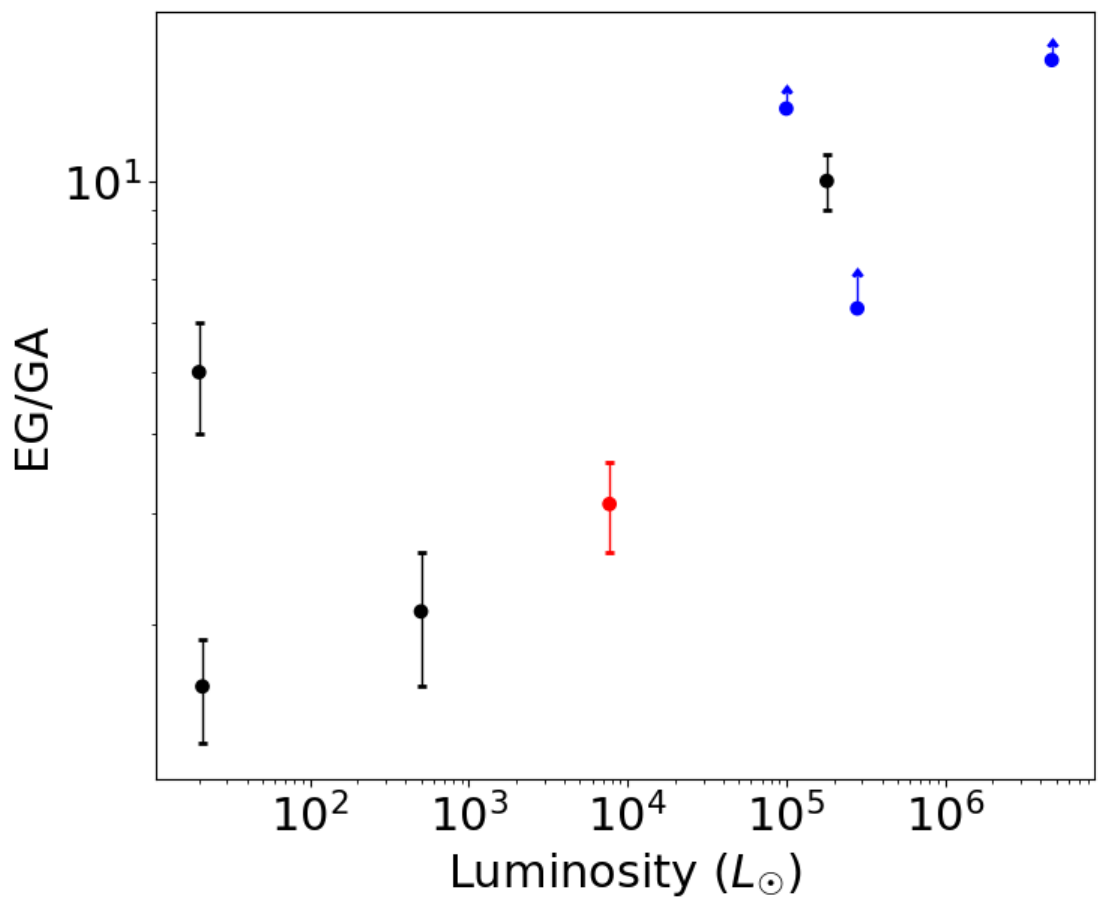}
\caption{Variation of the EG/GA ratios with luminosities in different high- and low-mass star-forming regions. The red circle is the EG/GA ratio of G358.93--0.03 MM1, and the blue circles are lower limits of EG/GA ratios towards Orion KL, G34.3+0.2, and W51e2.}
\label{fig:corelation}
\end{figure}

We compared the EG/GA ratio in G358.93--0.03 MM1 with other regions such as the massive star-forming regions Orion KL \citep{bro15}, W51e2 \citep{ly15}, G34.3+0.2 \citep{ly15}, the intermediate-mass hot core NGC 7129 FIRS2 \citep{fu14}, the HMC candidate G31.41+0.31 \citep{riv17}, the low-mass protostars IRAS 16293 B \citep{jo16}, and NGC 1333 IRAS 2A \citep{cou15}. The EG/GA ratios of the above-mentioned sources are shown in Tab.~\ref{tab:ratio}. The uncertainty of the EG/GA ratio in G358.93--0.03 MM1 can be deduced by considering the propagation of the uncertainty of the column densities of EG and GA. The behaviour of the EG/GA with respect to the luminosity of the different star-forming regions is shown in Fig.~\ref{fig:corelation}. As per Fig.~\ref{fig:corelation}, the EG/GA ratios increased with luminosities, with a lower value of 1 for the low-mass protostar IRAS 16293 B and a higher value of 10 for the HMC object G31.41+0.31. The lower limit of the hot cores in Orion KL, G34.3+0.2, and W51e2 is as high as 15. Earlier \citet{riv17} also see the same nature from the variation of the EG/GA ratios with luminosities in the star-forming regions, but they do not describe why these ratios increase with luminosities. We observe different EG/GA ratios because different initial compositions of the ices could produce very different values of the EG/GA ratios \citep{ob09}. Pure \ce{CH3OH} ices create a ratio of $>$10, but ices with a composition of \ce{CH3OH}:CO 1:10 produce a ratio of $<$0.25. This implies that the initial composition of the grains is significantly different for the high- and low-mass star-forming regions. Earlier, \cite{cou18} computed the two-phase warm-up chemical modelling of EG and GA, and they also observed that the modelled EG/GA ratios decreased with luminosities for different low- and high-mass star-formation regions (see Fig. 6 and 7 in \cite{cou18}). Subsequently, \cite{cou18} revised the physical assumptions in chemical modelling and added the grain mantle desorption. After this correction, they observed that the modelled EG/GA ratios increased with the luminosities (see Fig. ~12 in \cite{cou18}). We see that the modelled EG/GA ratios in \cite{cou18} cannot match any sources because the modelled EG/GA ratios with respect to luminosities are larger than the observed EG/GA ratios for different high- and low-mass sources.

\subsection{Possible formation pathways of EG in HMCs}
In HMCs, grain surface chemistry is crucial for the formation of COMs, including EG and GA. Previously, several grain surface chemical routes have been proposed, in which some of the reactions involved thermal and energetic processes \citep{gar08, bel09, wo12, gar13, but15}. In HMCs and hot corinos, the warm-up phase is thought to be important for the production of COMs. The warm-up phase allows the more strongly bound radicals to become mobile at the grain surface \citep{gar08, gar13}. Earlier, several experimental studies showed that the photochemistry in \ce{CH3OH}--CO ice mixtures is sufficient to explain the observed abundance of EG in the star-forming regions \citep{ob09}. Initially, two main formation pathways were proposed for the formation of EG\\\\
HCO + \ce{CH2OH} $\rightarrow$ \ce{CH2OHCHO} ~~~~~~~~~~~~(1)\\\\
\ce{CH2OHCHO} + H $\rightarrow$ \ce{CH2OHCHOH}~~~~~~~~(2)\\\\
\ce{CH2OHCHOH} + H $\rightarrow$ (\ce{CH2OH})$_{2}$~~~~~~~~~~~~(3)\\\\
and\\\\
HCO + 2H $\rightarrow$ \ce{CH2OH}~~~~~~~~~~~~~~~~~~~~~~~~~~~~~~(4)\\\\
\ce{CH2OH} + \ce{CH2OH} $\rightarrow$ (\ce{CH2OH})$_{2}$~~~~~~~~~~~~(5)\\\\
Reaction 1 indicates that the reaction between radical HCO and radical \ce{CH2OH} on the gain surfaces creates GA. Similarly, reactions 2 and 3 indicate that the subsequential hydrogenation of GA on the grain surface produces the EG \citep{fed15, ch16, gar13, cou18}. Earlier, \citet{riv17} raises many questions about the formation of EG using reactions 2 and 3. If EG is created via GA, a relatively constant EG/GA ratio would be expected \citep{riv17}. In Fig.~\ref{fig:corelation}, we observe that the EG/GA ratio varies by more than an order of magnitude (1 to $>$15) with increasing luminosity, which suggests that there is no direct link between EG and GA. The large variation in the EG/GA ratio indicates that GA is not a direct precursor of EG, as well as that they do not share a common precursor \citep{riv17}. We observe that EG is more abundant than GA in G358.93--0.03 MM1, which means the unsaturated molecule GA is broken towards the G358.93--0.03 MM1. For that reason, \cite{riv17} claims that reactions 2 and 3 are not sufficient to produce EG from GA. Earlier, \cite{req08} claimed that the double bond C = O in the unsaturated molecule GA can be easily broken in the gas phase of star-forming regions which indicates there is a possibility that GA can survive on the grain surface and EG can be produced by reactions 2 and 3. Similarly, reactions 4 and 5 indicate that the EG is formed by the recombination of two \ce{CH2OH} radicals in the grain surface, where radical \ce{CH2OH} is created via the subsequential hydrogenation of HCO \citep{gar08, gar13, but15, cou15, cou18, gar22}. \cite{mondal21} also used reaction 5 in the three-phase warm-up chemical models and showed that reaction 5 is the most efficient for the formation of EG in the HMC candidate G10.47+0.03. Previously, \cite{riv17} claimed that reaction 5 is the most efficient pathway to the formation of EG but \cite{cou18} showed that reaction 3 is the most efficient for the formation of EG towards the G31.41+0.31 using the two-phase warm-up chemical modelling. Recently, \cite{min23} supports the three-phase warm-up chemical model of \cite{gar22} and they show that reaction 5 is the most efficient route to the formation of EG towards the G31.41+0.31.

\subsection{Comparison between the modelled and observed abundance of EG}
Recently, \cite{gar22} developed a three-phase (gas + grain + ice mantle) warm-up chemical model using the three-phase astrochemical model code {\tt MAGICKAL} \citep{gar13} to understand the formation mechanisms and abundance of different COMs, including EG, towards the HMCs. To model a HMC, they first assumed a free-fall collapse of a cloud (phase I), followed by a warm-up phase (phase II). In the first phase, the gas density increases from $n_{H}$ = 3$\times$10$^{3}$ cm$^{-3}$ to 2$\times$10$^{8}$ cm$^{-3}$, and \cite{gar22} assume a constant temperature of 10 K. In the second phase (warm-up phase), the gas density remains constant at 2$\times$10$^{8}$ cm$^{-3}$ and the temperature increases with time from 8 to 400 K. \citet{gar22} used three different warm-up stages during the chemical model corresponding to a fast, medium, or slow warm-up from 8 to 400 K. The warm-up time scales of this chemical model were $5\times10^{4}$ yr (fast), $2\times10^{5}$ yr (medium), and $1\times10^{6}$ yr (slow). During chemical modelling, Garrod et al. (2022) used different types of chemical reactions for the formation of EG on the grain surface of HMCs (see Tab. 3 of \citet{gar22}). The estimated modelled abundances of EG with respect to \ce{H2} in \cite{gar22} was $6.0\times10^{-8}$, $1.1\times10^{-8}$, and $2.2\times10^{-12}$ corresponding to the fast, medium, and slow warm-up models (see Tab.~17 of \citet{gar22}). Similarly, the modelled abundances of EG with respect to \ce{CH3OH} was $5.5\times10^{-3}$, $1.0\times10^{-3}$, and $2.7\times10^{-7}$ corresponding to the fast, medium, and slow warm-up models (see Tab.~18 of \citet{gar22}). After chemical modelling, \cite{gar22} showed that the EG is formed via the recombination of two \ce{CH2OH} radicals (reaction 5) on the grain surface of HMCs.
 
To understand the formation pathways of EG towards the G358.93--0.03 MM1, we compare our estimated abundance of EG with the modelled abundance by \cite{gar22}. This comparison is physically reasonable because the maximum gas density and dust temperature of G358.93--0.03 MM1 are 1$\times$10$^{8}$ cm$^{-3}$ \citep{ste21} and $\sim$150 K \citep{chen20}, respectively. Hence, the three-phase warm-up chemical model of \cite{gar22} is appropriate for explaining the chemical evolution of EG towards the G358.93--0.03 MM1. We estimate the abundance of EG with respect to \ce{H2} towards the G358.93--0.03 MM1 to be (1.4$\pm$0.5)$\times$10$^{-8}$, which is very close to the medium warm-up model abundance of EG. Similarly, the abundance of EG with respect to \ce{CH3OH} towards the G358.93--0.03 MM1 is (6.1$\pm$0.3)$\times$10$^{-3}$, which is very close to the fast warm-up model abundance of EG with respect to \ce{CH3OH}. This comparison indicates that the simplest sugar alcohol molecule, EG, may form due to the recombination of two \ce{CH2OH} radicals (reaction 5) on the grain surface of G358.93--0.03 MM1. For G31.41+0.31, \cite{min23} estimated that the abundance of EG with respect to \ce{H2} is (1.5$\pm$0.4)$\times$10$^{-8}$, which is very close to the medium warm-up model abundance of EG. This means that reaction 5 may be responsible for the production of EG towards the G31.41+0.31. Earlier, \citet{en22} showed that reaction 5 has an activation barrier, and \citet{gar22} did not include an activation barrier during the chemical modelling. Therefore, it is necessary to verify if the main formation pathway of EG is via the recombination of two \ce{CH2OH} radicals by using a three-phase warm-up model with the activation barrier.

\section{Conclusion}
\label{conclu} 
We present the first confirmed detection of the rotational emission lines of the simplest sugar alcohol molecule, $aGg^{\prime}$-EG, towards the HMC candidate G358.93--0.03 MM1 using ALMA. The derived abundance of $aGg^{\prime}$-EG with respect to \ce{H2} towards the G358.93--0.03 MM1 is (1.4$\pm$0.5)$\times$10$^{-8}$. The EG/GA ratio towards the G358.93--0.03 MM1 is 3.1$\pm$0.5. Similarly, the abundance of $aGg^{\prime}$-EG with respect to \ce{CH3OH} towards the G358.93--0.03 MM1 is (6.1$\pm$0.3)$\times$10$^{-3}$. We observe that the fractional abundance of EG with respect to \ce{H2} towards the G358.93--0.03 MM1 is nearly similar to the abundance of EG towards the other two HMCs, G10.47+0.03 and G31.41+0.31 \citep{mondal21, min23}. We discuss the possible formation mechanisms of EG in HMCs. We compare our estimated abundance of EG with respect to \ce{H2} towards the G358.93--0.03 MM1 with the three-phase warm-up chemical model abundances of EG from \citet{gar22}, and we find that the derived and medium warm-up modelled abundances of EG are close. Similarly, we also compare the abundance of EG with respect to \ce{CH3OH} with the modelled abundance values of EG from \citet{gar22}, and we find that the derived and fast warm-up modelled abundances of EG with respect to \ce{CH3OH} are close. We also notice that the modelled abundance of EG by \citet{gar22} is similar to the observed abundance of EG towards other HMC candidate G31.41+0.31. So, EG may formed via the recombination of two \ce{CH2OH} radicals on the grain surface of G358.93--0.03 MM1 and G31.41+0.31. Earlier, \cite{en22} showed that the grain surface reaction between two \ce{CH2OH} radicals has an activation barrier, and \cite{gar22} did not include the activation barrier during the chemical modelling. As a result, it is essential to confirm whether the main formation mechanism of EG is via the recombination of two \ce{CH2OH} radicals, using a three-phase warm-up model that incorporates the activation barrier. The identification of high abundances of EG in G358.93--0.03 MM1 suggests that grain surface chemistry is efficient for the formation of other COMs, which should therefore be detectable. A combined spectral line study with an LTE model and three-phase warm-up chemical modelling is required to understand the hot core chemistry towards the G358.93--0.03 MM1, which will be carried out in our follow-up study.
	
\section*{ACKNOWLEDGEMENTS}{
We thank the anonymous reviewer and scientific editor Prof. Cecilia Ceccarelli for their helpful comments, which improved the manuscript. A.M. acknowledges the Swami Vivekananda Merit-cum-Means Scholarship (SVMCM) for financial support for this research. S.V.  acknowledges support from the European Research Council (ERC) under the European Union's Horizon 2020 research and innovation program MOPPEX 833460. This paper makes use of the following ALMA data: ADS /JAO.ALMA\#2019.1.00768.S. ALMA is a partnership of ESO (representing its member states), NSF (USA), and NINS (Japan), together with NRC (Canada), MOST and ASIAA (Taiwan), and KASI (Republic of Korea), in co-operation with the Republic of Chile. The Joint ALMA Observatory is operated by ESO, AUI/NRAO, and NAOJ.}
	
\section*{DATA AVAILABILITY}
The plots within this paper and other findings of this study are available from the corresponding author on reasonable request. The data used in this paper are available in the ALMA Science Archive (\url{https://almascience.nrao.edu/asax/}), under project code 2019.1.00768.S.

\bibliographystyle{aasjournal}

\begin{thebibliography}{}
	\expandafter\ifx\csname natexlab\endcsname\relax\def\natexlab#1{#1}\fi
	\providecommand{\url}[1]{\href{#1}{#1}}
	\providecommand{\dodoi}[1]{doi:~\href{http://doi.org/#1}{\nolinkurl{#1}}}
	\providecommand{\doeprint}[1]{\href{http://ascl.net/#1}{\nolinkurl{http://ascl.net/#1}}}
	\providecommand{\doarXiv}[1]{\href{https://arxiv.org/abs/#1}{\nolinkurl{https://arxiv.org/abs/#1}}}
	
\bibitem[\protect\citeauthoryear{Balucani et al.}{2015}]{ba15}Balucani, N., Ceccarelli, C., \& Taquet, V. 2015, MNRAS, 449, L16

\bibitem[\protect\citeauthoryear{Biver et al.}{2014}]{bi14}Biver, N., Bockelee-Morvan, D., Debout, V., et al. 2014, A\&A, 566, L5

\bibitem[\protect\citeauthoryear{Biver et al.}{2015}]{bi15}Biver, N., Bockelee-Morvan, D., Moreno, R., \& Crovisier, J. A. A. 2015, Sci. Adv., 1, e1500863

\bibitem[\protect\citeauthoryear{Brouillet et al.}{2015}]{bro15}Brouillet, N., Despois, D., Lu, X.-H., et al. 2015, A\&A, 576, A129

\bibitem[\protect\citeauthoryear{Bergner et al.}{2017}]{ber17}Bergner, J. B., {\"O}bergberg, K. I., Garrod, R. T., \& Graninger, D. M. 2017, ApJ, 841, 120

\bibitem[\protect\citeauthoryear{Bacmann et al.}{2012}]{bac12}Bacmann, A., Taquet, V., Faure, A., Kahane, C., \& Ceccarelli, C. 2012, A\&A, 541, L12

\bibitem[\protect\citeauthoryear{Belloche et al.}{2013}]{bel13}Belloche A., M\"uller H. S. P., Menten K. M., Schilke P., Comit C., 2013, A\&A, 559, A47

\bibitem[\protect\citeauthoryear{Beltr{\'a}n et al.}{2009}]{bel09}{Beltr{\'a}n} M.~T., Codella C., Viti S., Neri R., Cesaroni R., 2009, ApJL, 690, L93

\bibitem[\protect\citeauthoryear{Brogan et al.}{2019}]{bro19}Brogan C. L., Hunter T. R., Towner A. P. M. et al., 2019, ApJL, 881, L39

\bibitem[\protect\citeauthoryear{Bayandina et al.}{2022}]{bay22}Bayandina O.S., Brogan C.L., Burns, R.A., et al. 2022, AJ, 163(2), p.83

\bibitem[\protect\citeauthoryear{Butscher et al.}{2015}]{but15}Butscher T., Duvernay F., Theule P., Danger G., Carissan Y., Hagebaum-Reignier D., Chiavassa T., 2015, MNRAS, 453, 1587 

\bibitem[\protect\citeauthoryear{Blake et al.}{1987}]{bl87}Blake, G. A., Sutton, E. C., Masson, C. R., \& Phillips, T. G. 1987, ApJ, 315, 621

\bibitem[\protect\citeauthoryear{Cooper et al.}{2001}]{co01}Cooper, G., Kimmich, N., Belisle, W., et al. 2001, Nature, 414, 879

\bibitem[\protect\citeauthoryear{Christen et al.}{2001}]{ch01}Christen, D., Coudert, L. H., Larsson, J. A., \& Cremer, D. 2001, J. Mol. Spectr., 205, 185

\bibitem[\protect\citeauthoryear{Christen et al.}{1995}]{ch95}Christen, D., Coudert, L. H., Suenram, R. D., \& Lovas, F. J. 1995, J. Mol. Spectr., 172, 57

\bibitem[\protect\citeauthoryear{Chuang et al.}{2016}]{ch16}Chuang K. J., Fedoseev G., Ioppolo S., van Dishoeck E. F., Linnartz H., 2016, MNRAS, 455, 1702

\bibitem[\protect\citeauthoryear{Ceccarelli et al.}{2017}]{ce17}Ceccarelli, C., Caselli, P., Fontani, F., et al. 2017, ApJ, 850, 176

\bibitem[\protect\citeauthoryear{Ceccarelli et al.}{2023}]{ce23}Ceccarelli C. et al., 2023, ASP Conf. Ser., 534, 379

\bibitem[\protect\citeauthoryear{Coutens et al.}{2018}]{cou18}Coutens A., Viti S., Rawlings J.~M.~C., et al. 2018, MNRAS, 475, 2 

\bibitem[\protect\citeauthoryear{Coutens et al.}{2015}]{cou15}{Coutens} A., {Persson} M.~V., {J{\o}rgensen} J.~K., {Wampfler} S.~F., {Lykke} J.~M., 2015, A\&A, 576, A5

\bibitem[\protect\citeauthoryear{Chen et al.}{2020}]{chen20}Chen X., Sobolev A. M., Ren Z.-Y., et al. 2020, NatAs, 4, 1170

\bibitem[\protect\citeauthoryear{Crovisier et al.}{2004}]{cro04}Crovisier, J., Bockelee-Morvan, D., Biver, N., et al. 2004, A\&A, 418, L35

\bibitem[\protect\citeauthoryear{Enrique-Romero et al.}{2022}]{en22}Enrique-Romero J., Rimola A., Ceccarelli C., Ugliengo P., Balucani N., Skouteris D., 2022, ApJS, 259, 39

\bibitem[\protect\citeauthoryear{Fedoseev et al.}{2015}]{fed15}Fedoseev G., Cuppen H. M., Ioppolo S., Lamberts T., \& Linnartz H., 2015, MNRAS, 448, 1288

\bibitem[\protect\citeauthoryear{Favre et al.}{2018}]{fav18}Favre, C., Fedele, D., Semenov, D., et al. 2018, ApJ, 862, L2

\bibitem[\protect\citeauthoryear{Fayolle et al.}{2015}]{fa15}Fayolle, E. C., $\ddot{O}$berg, K. I., Garrod, R. T., van Dishoeck, E. F., \& Bisschop, S. E. 2015, A\&A, 576, A45

\bibitem[\protect\citeauthoryear{Favre et al.}{2011}]{fav11}Favre, C., Despois, D., Brouillet, N., et al. 2011, A\&A, 532, A32

\bibitem[\protect\citeauthoryear{Fuente et al.}{2014}]{fu14}Fuente, A., Cernicharo, J., Caselli, P., et al. 2014, A\&A, 568, A65

\bibitem[\protect\citeauthoryear{Gorai et al.}{2020}]{gor20}Gorai P., Bhat B., Sil M., Mondal S. K. Ghosh R., Chakrabarti S. K., Das A., 2020, ApJ, 895, 86

\bibitem[\protect\citeauthoryear{Garrod}{2013}]{gar13} Garrod R. T., 2013, ApJ, 765, 60	

\bibitem[\protect\citeauthoryear{Garrod \& Herbst}{2006}]{gar06}Garrod R. T., \& Herbst E. 2006., A\&A, 457, 927

\bibitem[\protect\citeauthoryear{Garrod et al.}{2008}]{gar08}Garrod R. T., Widicus Weaver S. L., \& Herbst E., 2008, ApJ, 682, 283

\bibitem[\protect\citeauthoryear{Garrod et al.}{2022}]{gar22}Garrod, R. T., Jin, M., Matis, K. A., Jones, D., Willis, E. R., Herbst, E., 2022, ApJs, 259, 1

\bibitem[\protect\citeauthoryear{Goesmann et al.}{2015}]{go15}Goesmann, F., Rosenbauer, H., Bredeh\"oft, J. H., et al. 2015, Science, 349, 020689

\bibitem[\protect\citeauthoryear{Hollis et al.}{2002}]{hol02}Hollis, J. M., Lovas, F. J., Jewell, P. R., \& Coudert, L. H. 2002, ApJ, 571, L59

\bibitem[\protect\citeauthoryear{Holdship et al.}{2017}]{hol17}Holdship J., Viti S., Jimenez-Serra I., Makrymallis A., Priestley F., 2017, AJ, 154, 38

\bibitem[\protect\citeauthoryear{Herbst \& van Dishoeck}{2009}]{her09}Herbst E., \& van Dishoeck E. F., 2009, ARA\&A, 47, 427

\bibitem[\protect\citeauthoryear{Isokoski et al.}{2013}]{is13}Isokoski, K., Bottinelli, S., \& van Dishoeck, E. F. 2013, A\&A, 554, A100

\bibitem[\protect\citeauthoryear{J{\o}rgensen et al.}{2016}]{jo16}{J{\o}rgensen} J. K. et al., 2016, A\&A, 595, A117

\bibitem[\protect\citeauthoryear{J{\o}rgensen et al.}{2020}]{jo20}J{\o}rgensen, J. K., Belloche, A., Garrod, R. T., 2020, ARA\&A, 58, 727

\bibitem[\protect\citeauthoryear{Lykke et al.}{2015}]{ly15}Lykke, J. M., Favre, C., Bergin, E. A., \& {J{\o}rgensen}, J. K. 2015, A\&A, 582, A64

\bibitem[\protect\citeauthoryear{Melosso et al.}{2020}]{mel20}Melosso, M., Dore, L., Tamassia, F., et al. 2020, JPCA, 124, 240

\bibitem[\protect\citeauthoryear{Manna \& Pal}{2023}]{man23a}Manna, A., \& Pal, S., 2023, Astrophys Space Sci 368, 33, 664

\bibitem[\protect\citeauthoryear{Manna et al.}{2023}]{man23b}Manna, A., Pal, S., Viti, S., Sinha, S., 2023, MNRAS, 525, 2229-2240

\bibitem[\protect\citeauthoryear{Manna \& Pal}{2024a}]{man24b}Manna A., \& Pal S., 2024a, New Astronomy, 109, 102199

\bibitem[\protect\citeauthoryear{Manna et al.}{2024}]{man24a}Manna, A., Pal, S., Baug, T., Mondal, S., 2024, Res. Astron. Astrophys, 24, 065008

\bibitem[\protect\citeauthoryear{Manna \& Pal}{2024b}]{man24c}Manna A., \& Pal S., 2024b, Res. Astron. Astrophys, 24, 075014

\bibitem[\protect\citeauthoryear{Mondal et al.}{2021}]{mondal21}Mondal S. K., Gorai P., Sil M., Ghosh R., Etim E. E., Chakrabarti S. K., Shimonishi T., Nakatani N., Furuya K., Tan J. C., Das, A., 2021, {ApJ}, 922, 194

\bibitem[\protect\citeauthoryear{McMullin et al.}{2007}]{mc07}McMullin J. P., Waters B., Schiebel D., Young W., \& Golap K. 2007, in Astronomical Society of the Pacific Conference Series, Vol. 376, Astronomical Data Analysis Software and Systems XVI, ed. R. A. Shaw, F. Hill, \& D. J. Bell, 127

\bibitem[\protect\citeauthoryear{M\"uller et al.}{2005}]{mu05} M\"uller H. S. P., SchlM$\ddot{o}$der F., Stutzki J. Winnewisser G., 2005, Journal of Molecular Structure, 742, 215

\bibitem[\protect\citeauthoryear{M\"uller \& Christen}{2004}]{mu04}M\"uller, H. S. P., \& Christen, D. 2004, J. Mol. Spectr., 228, 298

\bibitem[\protect\citeauthoryear{Mininni et al.}{2023}]{min23}Mininni, C., Beltr{\'a}n, M.~T., Colzi, L., et al. 2023, A\&A, 677, A15

\bibitem[\protect\citeauthoryear{Ospina-Zamudio et al.}{2018}]{os18}Ospina-Zamudio, J., Lefloch, B., Ceccarelli, C., et al. 2018, A\&A, 618, A145

\bibitem[\protect\citeauthoryear{{\"O}berg et al.}{2009}]{ob09}{\"O}berg, K. I., Garrod, R. T., van Dishoeck, E. F., \& Linnartz, H. 2009, A\&A, 504, 891

\bibitem[\protect\citeauthoryear{Perley \& Butler}{2017}]{pal17}Perley R. A., Butler B. J., 2017, ApJ, 230, 1538

\bibitem[\protect\citeauthoryear{Puzzarini}{2022}]{puz22}Puzzarini C. 2022, FrASS, 8, 811342

\bibitem[\protect\citeauthoryear{Reid et al.}{2014}]{re14}Reid M. J., Menten K. M., Brunthaler, A., et al. 2014, ApJ, 783, 130	

\bibitem[\protect\citeauthoryear{Rivilla et al.}{2017}]{riv17}Rivilla V. M., Beltr{\'a}n M. T., Cesaroni R., et al. 2017, A\&A, 598, A59

\bibitem[\protect\citeauthoryear{Requena-Torres et al.}{2008}]{req08}Requena-Torres, M. A., Martin-Pintado, J., Martin, S., \& Morris, M. R. 2008, ApJ, 672, 352

\bibitem[\protect\citeauthoryear{Shimonishi et al.}{2021}]{shi21} Shimonishi T., Izumi N., Furuya K., \& Yasui C., 2021, ApJ, 2, 206

\bibitem[\protect\citeauthoryear{Stecklum et al.}{2021}]{ste21}Stecklum B., Wolf V., Linz H., et al., 2021. A\&A, 646, p.A161

\bibitem[\protect\citeauthoryear{Skouteris et al.}{2018}]{sk18}Skouteris, D., Balucani, N., Ceccarelli, C., et al. 2018, ApJ, 854, 135

\bibitem[\protect\citeauthoryear{Taquet et al.}{2017}]{taq17}Taquet, V., Wirstr$\ddot{o}$m, E. S., Charnley, S. B., et al. 2017, A\&A, 607, A20

\bibitem[\protect\citeauthoryear{Vastel et al.}{2015}]{vas15}Vastel C., Bottinelli S., Caux E., Glorian J. -M., Boiziot M., 2015, CASSIS: a tool to visualize and analyse instrumental and synthetic spectra. Proceedings of the Annual meeting of the French Society of Astronomy and Astrophysics, 313-316

\bibitem[\protect\citeauthoryear{Viti et al.}{2004}]{vi04}Viti, S., Collings, M. P., Dever, J. W., McCoustra, M. R. S., \& Williams, D. A. 2004, MNRAS, 354, 1141

\bibitem[\protect\citeauthoryear{van Dishoeck \& Blake}{1998}]{van98}van Dishoeck E. F. \& Blake G. A., 1998, Annu Rev Astron Astrophys, 36, 317

\bibitem[\protect\citeauthoryear{Woods et al.}{2012}]{wo12}Woods P. M., Kelly, G., Viti, S., et al., 2012, APJ, 750, 19

\bibitem[\protect\citeauthoryear{Williams \& Viti}{2014}]{wi14}Williams D. A., Viti S., 2014, Observational Molecular Astronomy: Exploring the Universe Using Molecular Line Emissions. Cambridge Univ. Press, Cambridge

\bibitem[\protect\citeauthoryear{Walsh et al.}{2016}]{wal16}Walsh, C., Juh$\acute{a}$sz, A., Meeus, G., et al. 2016, ApJ, 831, 200	

\end{thebibliography}

% Don't change these lines
\bsp	% typesetting comment
\label{lastpage}
\end{document}